\documentclass[aps,prx,twocolumn,superscriptaddress,notitlepage]{revtex4-1}

\usepackage{graphicx}
\usepackage{epsfig}
\usepackage{epstopdf}
\usepackage{float}
\usepackage{subfigure}
\usepackage{bbold}

\usepackage{dcolumn}
\usepackage{latexsym}
\usepackage{amssymb}
\usepackage{amsmath}
\usepackage{amsfonts}
\usepackage{wasysym}
\usepackage{bm}

\usepackage[colorlinks,bookmarks=false,citecolor=blue,linkcolor=red,urlcolor=blue]{hyperref}
\usepackage{verbatim}

\usepackage{booktabs}

\usepackage{algorithm}
\usepackage{algorithmic}
\usepackage{indentfirst}
\usepackage{listings}

\newcommand{\ud}{\mathrm{d}}
\newcommand{\oh}{\frac{1}{2}}

\newcommand{\lefta}{\left\langle}
\newcommand{\righta}{\right\rangle}

\newcommand{\nn}{\nonumber}

\newcommand{\RNum}[1]{\uppercase\expandafter{\romannumeral #1\relax}}
\newcommand{\tx}[1]{\textmd{#1}}
\newcommand{\onefrac}[1]{\frac{1}{#1}}
\newcommand{\refeq}[1]{\textmd{Eq.\ }(\ref{#1})}
\newcommand{\reffg}[1]{\textmd{Fig.\ }\ref{#1}}
\newcommand{\refapp}[1]{\textmd{Appendix\ }\ref{#1}}

\DeclareMathOperator{\Tr}{Tr}

\newcommand{\im}{\operatorname{Im}}
\newcommand{\re}{\operatorname{Re}}

\newcommand{\vrr}{\mathbf{r}}
\newcommand{\vq}{\mathbf{q}}

\newcommand{\vk}{\mathbf{k}}

\def\be{\begin{equation}}
\def\ee{\end{equation}}
\def\bea{\begin{eqnarray}}
\def\eea{\end{eqnarray}}

\usepackage{array}
\newcommand{\PreserveBackslash}[1]{\let\temp=\\#1\let\\=\temp}
\newcolumntype{C}[1]{>{\PreserveBackslash\centering}p{#1}}
\newcolumntype{R}[1]{>{\PreserveBackslash\raggedleft}p{#1}}
\newcolumntype{L}[1]{>{\PreserveBackslash\raggedright}p{#1}}

\newcommand{\ve}[1]{{\bf #1}}
\newcommand{\kv}{\ve{k}}
\newcommand{\qv}{\ve{q}}

\usepackage{color}

\definecolor{darkblue}{rgb}{0,0.02,0.45}
\definecolor{darkred}{rgb}{0.45,0.02,0} 

\makeatletter
\newcommand{\thickhline}{%
\noalign {\ifnum 0=`}\fi \hrule height 0.7pt
\futurelet \reserved@a \@xhline
}
\newcolumntype{"}{@{\hskip\tabcolsep\vrrule width 0.7pt\hskip\tabcolsep}}
\makeatother

\usepackage{times}

\begin{document}

\title{Temperature evolution of the phonon dynamics in the Kitaev spin liquid}

\author{Kexin Feng}
\affiliation{School of Physics and Astronomy, University of Minnesota, Minneapolis, MN 55455, USA}
\author{Mengxing Ye}
\affiliation{Kavli Institute for Theoretical Physics, University of California, Santa Barbara, CA 93106, USA}
\author{Natalia B. Perkins}
\affiliation{School of Physics and Astronomy, University of Minnesota, Minneapolis, MN 55455, USA}
\date{\today}

\begin{abstract}
  Here we  present a study  of the phonon dynamics in the honeycomb Kitaev spin model at finite temperatures.
  We show that the fractionalized spin excitations of the Kitaev spin liquid, the itinerant Majorana fermions and static $Z_2$ fluxes, have distinct effects on the phonon dynamics, which makes the phonon dynamics a promising tool for exploring the Kitaev spin liquid candidate materials.
In particular, we focus on the signature of the fractionalized excitations in the thermodynamic behaviour of the sound attenuation and the phonon Hall viscosity:
The former describes the phonon decay into the fractionalized excitations, and the latter is the leading order time reversal symmetry breaking effect on the acoustic phonon.
 We  find that the angular dependence of the attenuation coefficient  and its magnitude are modified by the thermal excitation of the  $Z_2$ fluxes. The strength of this effect strongly depends  on  the relative  magnitude of the sound velocity and the Fermi velocity characterizing the low-energy Majorana fermions.
 We also show that the Hall viscosity is strongly suppressed by the increase of the density of the $Z_2$ fluxes at finite temperatures.
 All these  observations reflect the effects of the 
 emergent disorder on the Majorana fermions introduced by the $Z_2$ fluxes. 
 Our analysis is based on the complementary analytical calculations in the low-temperature zero-flux sector, and numerical 
 calculations in the inhomogeneous flux sectors at intermediate and high temperatures with stratified Monte Carlo (strMC) method.
\end{abstract}

\maketitle

\section{Introduction}
Recent years have seen a lot of progress in identifying candidate materials that can realize  quantum spin liquid (QSL) phases  ~\cite{Anderson1973}, the interest to which stems from their
 remarkable set of collective phenomena including topological ground-state degeneracy, long-range entanglement, and fractionalized excitations \cite{Wen2002,Kitaev2006,Balents2010,Savary2016,KnolleMoessner2019,Broholm2020}. 
In particular, a significant experimental and theoretical effort has been devoted to the study of magnetic properties of spin-orbit coupled 4d and 5d magnets, dubbed Kitaev materials~\cite{Jackeli2009,Trebst2017,Takagi2019,Motome2019}, which can potentially realize the celebrated Kitaev honeycomb model \cite{Kitaev2006}.

The experimental detection and characterization of the  QSL states  are, however, very difficult since
the absence of local magnetic orders down to zero temperatures makes them in many aspects invisible  to the conventional local  probes.  
When searching for QSL physics in the  Kitaev materials,  a promising route  is to look for signatures of spin fractionalization into two types of quasiparticle excitations,  which according to the exact solution of the Kitaev model  are  localized,  gapped $Z_2$ fluxes and itinerant,  gapless Majorana fermions.  The dynamical probes which have been heavily exploited for this goal are
    inelastic neutron scattering \cite {Knolle2015,Banerjee2016,Banerjee2017,banerjee2018excitations},  Raman scattering \cite{Knolle2014b,Sandilands2015,Perreault2015,Perreault2016,Rousochatzakis2019,Sahasrabudhe2020,Yiping2020,Dirk2020}, resonant inelastic x-ray scattering \cite{Gabor2016,Gabor2017,Gabor2019,Ruiz2021}, ultrafast spectroscopy \cite{Alpichshev15,Zhang2019}  and 2D terahertz non-linear coherent spectroscopy \cite{Little2017,Wan2019}. 

The search for fractionalization in QSLs largely relies on their decoupling with the environment. However, the spin-lattice interaction is inevitable in real  materials and  is often rather strong \cite{Hentrich2018,Kasahara2018,Pal2020,Miao2020}.  In particular,  
since the dynamics of phonons can be modified by 
the coupling of phonons to fractionalized excitations \cite{Metavitsiadis2020,Ye2020},  
 the study of  phonon dynamics  can  serve as an additional probe of fractionalization and provide important information about the nature of the QSL state. In the Kitaev spin liquid, the coupling of phonons to Majorana fermions is also expected to play a significant role in the thermal Hall effect \cite{Ye2018b,Rosch2018}, the observation of which in the Kitaev material $\ensuremath{\alpha}\text{\ensuremath{-}}{\mathrm{RuCl}}_{3}$ \cite{Kasahara2018}, 
 considered to be proximate  to the Kitaev QSL state, is one of the most direct evidences of the presence of fractional excitations in this material.
 
Recently, some of us have shown that the phonon dynamics may serve as an indirect probe of  fractionalization of  spin degrees of freedom in the Kitaev QSL through the study of the sound attenuation ($\alpha_s$) from the phonon  decaying into a pair of Majorana fermions, and the Hall viscosity ($\eta_H$) induced by the time-reversal breaking spin Hamiltonian \cite{Ye2020}. Specifically, the sound attenuation may be measured by the ultrasound experiment~\cite{Pippard1955,Akhiezer1957,Blount1959,Tsuneto1961,Ott1985,Kazumi1994}, and the Hall viscosity could be inferred from the acoustic Faraday effect~\cite{Boiteux1971,Sytcheva2010,Tuegel2017}, thermal Hall effect~\cite{Shi2012,Rosch2018,Xiao2019,Chen2020,Ye2020,Ye2021a} and spectroscopy measurement.  In order to compute these quantities,  in Ref. \cite{Ye2020} we derived a low-energy effective spin-lattice coupling in terms of the matter Majorana fermions and the acoustic phonons and formulated a diagrammatic computation procedure.  As we were focusing only on the low-temperature  response, the flux degrees of freedom were totally neglected.
However, since thermal excitations  of fluxes at finite temperatures  significantly modify the Majorana fermion spectrum, they can give  rise  to  dynamical responses  that  are  strikingly different from  their zero-temperature counterparts \cite{Gabor2019,Rousochatzakis2019,Metavitsiadis2020}.


In this work, we study the phonon dynamics at finite temperatures and focus on the effects of the $Z_2$ fluxes on the temperature dependence of the sound attenuation and Hall viscosity coefficients. Importantly, as the $Z_2$ flux excitations are static and do not couple to phonons directly~\footnote{This statement assumes that the spin-lattice couplings are restricted to the Kitaev interaction form, i.e.\ it only modifies the strength of the spin interaction but does not generate additional forms.}, the renormalized phonon propagator can still be obtained  from the polarization bubble whose internal states are just Majorana fermions, and the problem  is still exactly solvable.
However, in order to compute this polarization bubble in the presence of the fluxes
when the translational symmetry for the Majorana fermions is broken, we had to derive a microscopic low-energy effective spin-lattice coupling   and  a  diagrammatic  computation procedure in the mixed representation when the Majorana fermion eigenmodes  are obtained from the diagonalization in the real space  and the acoustic phonons are treated in the momentum space.

We show that the sound attenuation 
manifests a clear six-fold symmetry with  respect to the 
phonon momentum both in the zero-flux sector and when the sound attenuation coefficient is averaged over different thermal flux sectors.
Second, the sound attenuation
 in the presence of fluxes shows  a very different temperature evolution 
 depending on the relative  magnitude of the sound velocity and the  Fermi velocity characterizing the low-energy Majoranas.

When the sound velocity is smaller than the Fermi velocity,  the sound attenuation
 at low-temperature zero-flux sector is determined by the microscopic processes in which  a Majorana fermion
 is  excited to a higher energy fermion state (dubbed as ph-channel as we define  in Sec.~\ref{subsec: basic_picture}), with the attenuation rate linear in temperature due to the vanishing density of states at the Dirac points \cite{Ye2020}.  These processes satisfy both the energy and momentum kinematic constraints.   
 At intermediate and high temperatures, the relaxation of the momentum kinematic constraint and the modification of the fermionic spectrum due to the thermal flux lead to a 
 significant change in the angular distribution of the attenuation coefficient in the momentum space
and
an overall decrease of the magnitude of the phonon decay.

When the sound velocity is larger than the Fermi velocity, in the  low-temperature zero-flux sector the kinematic constraints   can only  be satisfied in the microscopic processes when a  phonon decays into  two fermions, both with positive energy (dubbed as pp-channel in Sec.~\ref{subsec: basic_picture}). These processes can happen even  at  zero temperature as long as a phonon has enough energy to excite a pair of particles.    Consequently, both the intensity and the angular pattern of the attenuation coefficient  do not strongly depend on 
the temperature.

In the presence of the time-reversal symmetry breaking term in the spin Hamiltonian, we study  its effect (in the leading order)  on the phonon dynamics through a phonon Hall viscosity~\cite{avron1995viscosity,barkeshli2012dissipationless}. 
Our results  on its  temperature dependence show that while at 
low  temperatures, when the zero-flux is a good approximation of the thermal flux, the phonon Hall viscosity remains almost temperature independent,  at  higher temperatures
 there is a significant reduction of the phonon Hall viscosity coefficient due to the presence of the $Z_2$ fluxes. We also show that, counter-intuitively,  the phonon Hall viscosity  exhibits a sizeable decrease  as the strength of the time-reversal symmetry breaking perturbation in the spin Hamiltonian increases.
 
 In  order to study the phonon dynamics above the flux proliferation temperatures, we developed a  method to sample the flux configurations,  dubbed stratified Monte Carlo (strMC) algorithm.  This algorithm stratifies the sample space according to the flux pseudo-potential energy  model proposed in Ref.\ \cite{feng2020} and yields unbiased results of the thermodynamic quantities calculated throughout this paper. This algorithm has several advantages over the commonly used Markov Chain Monte Carlo (MCMC) algorithm. It fundamentally reduces the autocorrelation time to zero and accelerates the convergence. It is also convenient in implementation and parallelization, and free from the local minima problem.
The details of this algorithm is presented in \refapp{app: sMC}.

 The  rest of  the  paper  is  organized  as  follows. In Section ~\ref{Sec:SPH}, we introduce the  general  aspects  of  the   spin-phonon Hamiltonian. In
  Section~\ref{sec:RSF}, we show how to obtain the finite-temperature  fermionic spectrum  by solving the free-fermion problem  in each flux sector exactly.
  We show that  in the presence of fluxes the diagonalization of the Majorana fermion Hamiltonian should be performed  in the real space and  can be done by using singular value decomposition. Then, we introduce the imaginary time fermion propagators  which we will later use in the calculations of observable quantities.  In Section~\ref{sec:MFPh},  we derive the  Majorana fermion-phonon (MFPh) coupling in the  mixed representation.
 We then proceed to Section~\ref{sec:III}  and  discuss  the effect of the spin-lattice coupling on the  phonon dynamics. Here we compute  the phonon  polarization bubble, which is the  key quantity defining the phonon dynamics.  
In Section~\ref{sec:IV},  we relate the imaginary part of the diagonal components of the phonon polarization bubble to the attenuation coefficients. 
 In Section~\ref{subsec: basic_picture}, we discuss the role  of the kinematic constraints on the phonon dynamics.
In Section~\ref{subsec:IVB}, we present the analysis of  the phonon dynamics in the  thermal flux sectors, when the kinematic constraint of momentum conservation is relaxed.    Based on this analysis,  in Section~\ref{subsec:IVC} we  present  numerical results for the temperature evolution of the  sound attenuation coefficient   obtained by employing the stratified Monte Carlo (strMC) algorithm.
The details of the strMC method, the comparison with MCMC method, and its tailored application to the Kitaev honeycomb model are discussed in \refapp{app: sMC}.
 In Section~\ref{sec:V},  we  study the observable consequences of the  spin-lattice coupling  when   time reversal symmetry  is broken. In particular, we study the temperature evolution of the Hall viscosity coefficient, $\eta_H$ 
focusing on the understanding the effect of the $Z_2$ fluxes.
We conclude with a general discussion of our results in Section~\ref{sum}. 
 Technical details and other auxiliary information are provided in Appendices~\ref{app: eval_bubble}-\ref{sec: MCkap}.

\section{The model} \label{Sec:model}
\subsection{The spin-phonon Hamiltonian} \label{Sec:SPH}

We focus our discussion on the spin-phonon Hamiltonian:
\begin{align}
\mathcal{H}=\mathcal{H}_{ s}+\mathcal{H}_{\mathrm{ph}}+\mathcal{H}_{\text c}, \label{eq:model}
\end{align} 
For simplicity, we consider the isotropic Kitaev model on a two-dimensional (2d) honeycomb lattice. The model has $G=P6mm$ space group
(group number $77$), which is a semidirect product of the point group ${\rm C}_{6v}$~\cite{you2012doping} and translation group $P$ on a 2d hexagonal lattice, i.e.\ $G={\rm C}_{6v}\ltimes P$. The space group symmetry gives constraints on the form of $\mathcal{H}$, which we discuss in detail below. Importantly, while a given flux configuration breaks the space group symmetry, the symmetry is restored after averaging over all flux configurations due to thermal fluctuations.

\begin{figure}
	\centering
	\includegraphics[width=1.0\columnwidth]{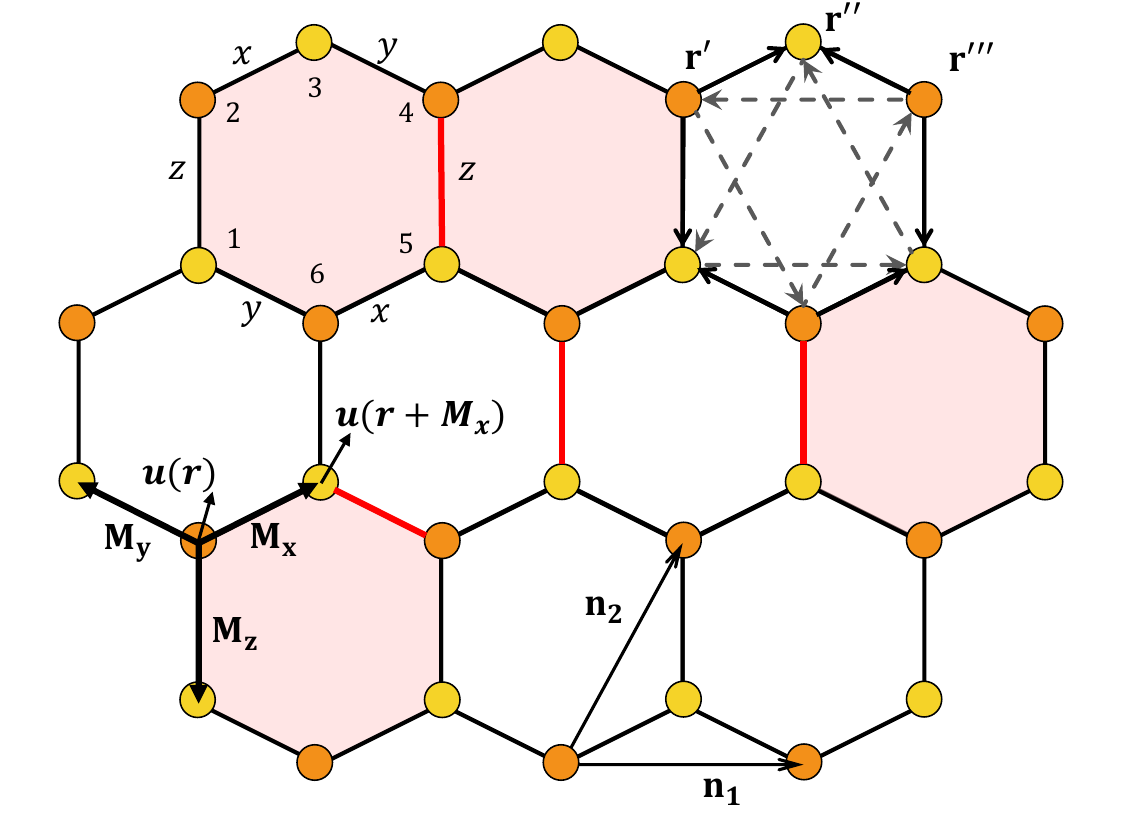}
     \caption{Kitaev model on the honeycomb lattice with unit vectors ${\bf n}_1=(1,0)$ and  ${\bf n}_2=(\frac{1}{2},\frac{\sqrt{3}}{2})$ shown in a random thermal flux sector. The sites on the  A and B sublattices of the honeycomb lattice  are marked by yellow and orange circles. The three vectors 
     $\mathbf{M}_{x, y}=\left(\pm \frac{\sqrt{3}}{2}, \frac{1}{2}\right),$ and $\mathbf{M}_{z}=(0,-1)$ connect nearest neighbors (NN) by $x, y,$ and $z$ bonds, respectively.  The black arrow represent the  displacement  vector ${\bf u} ({\bf r})=\{u_x,u_y\}$.
The red hexagon on the top left plaquette shows a $\pi$ flux, i.e. the eigenvalue of  $\hat{W}_p = \sigma_1^x \sigma_2^y \sigma_3^z \sigma_4^x \sigma_5^y \sigma_6^z$ is equal to -1. The red thick lines represent the link gauge variable with $\eta_{lm}=-1$, which creates or annihilates fluxes on its two adjacent plaquettes. The arrows on the top right plaquette show the convention of the signs of the NN couplings and NNN couplings, e.g.\ an arrow pointing from ${\bf r'}$ to ${\bf r}''$ means that the value of the NN bond $\eta_{ {\bf r'}{\bf r}''}$ is by default positive in \refeq{eq:Hamfermion}.
The sites $\langle {\bf r}'{\bf r}''{\bf r}'''\rangle$ is an example of the NNN triplet used in \refeq{eq:Kmodel1} and corresponds to $\lefta lnm \righta$ in \refeq{eq:Hamfermion}.
}
\label{Fig:lattice}
\end{figure}

The {\it first term}  in \refeq{eq:model} represents the spin  Hamiltonian given by
\begin{align}
\mathcal{H}_{\text s}  = -\sum_{\alpha,{\bf r}\in A}  J^\alpha  \sigma_{\bf r}^{\alpha} \sigma_{{\bf r}+{\bf M}_\alpha}^{\alpha} - 
\kappa \sum_{\langle {\bf r}',{\bf r}'',{\bf r}''' \rangle_{\alpha\gamma}}\sigma^\alpha_{{\bf r}'} \sigma^\beta_{{\bf r}''} \sigma^\gamma_{{\bf r}'''}. \label{eq:Kmodel1}
\end{align}
where $J^{\alpha}$  denotes the nearest neighbor Kitaev interaction  on  the corresponding bond of type $\alpha=x,y,z$,  $\sigma^\alpha_{\bf r}$ are the Pauli matrices, and ${\bf M}_\alpha$ labels the three inequivalent bonds on the honeycomb lattice (see \reffg{Fig:lattice}). In the isotropic Kitaev model, we consider $J^\alpha=J_K$.
The second term in  Eq.(\ref{eq:Kmodel1}) breaks time-reversal and vertical mirror symmetries, but preserves the exact solubility of the model. It mimics the effect of an external magnetic field ${\bf h}=(h_x,h_y,h_z)=\frac{h}{\sqrt{3}}(1,1,1)$ perturbatively, with $\kappa \sim \frac{h_{x}h_{y}h_{z}}{J^{2}}$.
 The three-spin link notation $\langle {\bf r}',{\bf r}'',{\bf r}''' \rangle_{\alpha\gamma}$ shown in \reffg{Fig:lattice} labels bonds $\bf {r}'\bf {r}''$, $\bf {r}''\bf {r}'''$ of type $\alpha,\gamma$, respectively, on three adjacent sites ${\bf r}',{\bf r}'',{\bf r}'''$.

The {\it second term} in  Eq.(\ref{eq:model})  is the bare Hamiltonian for the acoustic phonons on the honeycomb lattice, 
 \begin{align}
    \mathcal{H}_{\mathrm{ph}}&=\mathcal{H}^{kinetic}_{\mathrm{ph}}+\mathcal{H}^{elastic}_{\mathrm{ph}}.
 \end{align}
Here,
 $\mathcal{H}^{kinetic}_{\mathrm{ph}}=\sum_{{\bf q}}\frac{\bf {P}_{-\bf q}\cdot \bf {P}_{\bf q}}{2\rho\delta_V}$,  
where $\bf {P}_{{\bf q}}=\rho \,\partial_t {\bf u}_{\bf q}$ is the momentum  of the phonon. $\delta_V$ is the area enclosed in one unit cell, $\rho$ is the mass density of the lattice ion, and ${\bf u}=\{u_x,u_y\}$ is the lattice displacement vector. The elastic energy is determined by the point group of the crystal. For the ${\rm C}_{6v}$ group of our interest, it is expressed in terms of the strain tensor $\epsilon_{\alpha\beta}=\frac{1}{2}(\partial_\alpha u_\beta+\partial_\beta u_\alpha)$, as
\begin{align}
    \mathcal{H}^{elastic}_{\mathrm{ph}}=& \int d^2 x\, \big[C_1 (\epsilon_{xx}+\epsilon_{yy})^2\nonumber\\
&+ C_2(\epsilon_{xx}-\epsilon_{yy}+2i\epsilon_{xy})(\epsilon_{xx}-\epsilon_{yy}-2i\epsilon_{xy})\big]. \label{eq:Phmodel1}
\end{align}
Here, $\epsilon_{xx}+\epsilon_{yy}$ and $\{\epsilon_{xx}-\epsilon_{yy}, 2\epsilon_{xy}\}$ are from the basis of $A_1^{ph}$ and $E_2^{ph}$ irreducible representations (IRRs) of the ${\rm C}_{6v}$ point group, respectively, and $C_1$ and $C_2$ are the only two independent nonzero elastic module tensor coefficients \cite{Ye2020}. 
The lattice symmetry holds even in the presence of fluxes, so we can still use symmetry considerations to describe phonon modes at finite temperatures.
Based on the analysis of Ref. \cite{Ye2020},  we can write the longitudinal/transverse acoustic phonon spectrum and the polarization vectors (defined through ${\bf{u}}_{\bf q}=\sum_{\nu}\hat{e}_{\bf{q}}^{\nu}\tilde u_{\bf{q}}^{\nu}$)  as
\begin{align}
\Omega^{\parallel}_{\bf q}=v_s^{\parallel}q=\sqrt{\frac{C_1+C_2}{\rho}} \,q, \quad  \hat{e}_{\bf{q}}^{\parallel}=\{\cos\theta_\vq,\sin\theta_\vq\}\nonumber\\
\Omega^{\perp}_{\bf q}=v_s^{\perp}q=\sqrt{\frac{C_2}{\rho}} \,q,  \quad \hat{e}_{\bf{q}}^{\perp}=\{-\sin\theta_\vq,\cos\theta_\vq\},
\label{appeq:PhSpectrum}
\end{align}
where $q=\sqrt{q_x^2+q_y^2}$  and  $\theta_\vq$  is defined  as the angle between ${\bf q}$ and ${\hat x}$ axis,  

 Knowing the acoustic  phonon dispersion relations,  we can now determine the free phonon propagator in terms of lattice displacement field $\tilde{u}_{{\bf q},\nu}$ as
\begin{align}
D_{\nu\nu}^{(0)}({\bf q},t)=-i\langle T_t \tilde{u}_{-{\bf q}}^{\nu}(t)\tilde{u}_{{\bf q}}^{\nu}(0)\rangle^{(0)},
\label{BarePHpropagator}
\end{align}
where  $T_t$ is time ordering operator, $\nu=\|,\perp$ labels the polarization, and  the
  quantized displacement field is given by the standard expression: 
\begin{align}
    \tilde{u}_{\bf q}^{\nu}(t) &= i\sqrt{\frac{\hbar}{2 \rho \delta_{V} \Omega_{\bf q}^{\nu}}}\left(a_{{\bf q}, \nu} e^{-i \Omega_{\bf q}^{\nu}t}+a_{{-\bf q},\nu }^{\dagger} e^{i \Omega_{\bf q}^{\nu}t}\right).\label{eq: phonon_quanti}
\end{align}

In the rest of the discussions we set $\hbar=1$.  Also, since
 in the following we will focus on the finite-temperature physics,  it is convenient for us to  rewrite the phonon propagator using the Matsubara formalism \cite{Mahan}:
\begin{align}
    D^{(0)}_{\nu\nu}({\bf q}, i\Omega_n)  &= -\frac{1}{\rho \delta_{V}} \frac{1}{\left(i \Omega_{n}\right)^{2}-(\Omega_{\bf q}^{\nu})^{2}}. \label{eq: phonon_prop}
\end{align}

The {\it third term} term  in Eq.(\ref{eq:model}) denotes the magneto-elastic coupling that arises from the change in the Kitaev coupling $J_K$ due to the lattice  vibrations. In the long wavelength limit for the acoustic phonons, assuming that $J_K$ only depends on the distance $r$ between the atoms, the coupling Hamiltonian can be written in a  differential form:
\begin{eqnarray}
\label{Cmodel1}
\mathcal{H}_{\text c}  &&=- \lambda\sum_{{\bf r},\alpha}  {\bf M}_\alpha
\cdot \left(
{\bf u}({\bf r})- {\bf u}({\bf r}+{\bf M}_\alpha)
\right)  
\sigma_{\bf r}^{\alpha}  \sigma_{{\bf r}+{\bf M}_\alpha}^{\alpha}
\\\nonumber
&&= \lambda\sum_{{\bf r},\alpha}  {\bf M}_\alpha\cdot \left[ \left(
{\bf M}_\alpha \cdot{\bf \nabla} \right)
{\bf u}({\bf r})\right] \sigma_{\bf r}^{\alpha}  \sigma_{ {\bf r}+{\bf M}_\alpha }^{\alpha},
\end{eqnarray}
where $\lambda \sim\left(\frac{\mathrm{d} J_{K}}{\mathrm{d} {r}}\right)_{e q} l_{a}$  is the strength of the spin-phonon interaction and   $l_{a}$ is the lattice constant. Note that since the strength of the three-spin interaction term $\kappa$ depends $J_K$, it  will also change  due to the lattice  vibrations. However the contribution from this term to 
the  magneto-elastic coupling  will appear at higher orders in $h/J_K$, and thus will  be neglected.

The general symmetry allowed spin-lattice coupling can be expressed in terms of $A_1$ and $E_2$ channels 
independently~\cite{Ye2020}. In Ref.\cite{Ye2020}, it was shown that the dominant contribution to the magneto-elastic coupling comes from the $E_2$ symmetry channel, and in the following we will focus only on this channel \footnote{ We have checked  numerically that the contribution from the $A_1$ channel remains parametrically small  even in the presence of the $Z_2$ fluxes.}.
Then, the magnetoelastic coupling is given by
\begin{align}
    &\mathcal{H}_{\text c}^{E_{2}} = \lambda_{E_{2}} \sum_{\mathbf{r}\in A} \left[\ 2 \sqrt{3} \epsilon_{x y}\left(\sigma_{\mathbf{r}}^{x} \sigma_{\mathbf{r}+\mathbf{M}_{x}}^{x}-\sigma_{\mathbf{r}}^{y} \sigma_{\mathbf{r}+\mathbf{M}_{y}}^{y}\right) \right.\nn
    \\
    &
   +\left.\left(\epsilon_{x x}-\epsilon_{y y}\right)\left(\sigma_{\mathbf{r}}^{x} \sigma_{\mathbf{r}+\mathbf{M}_{x}}^{x}+\sigma_{\mathbf{r}}^{y} \sigma_{\mathbf{r}+\mathbf{M}_{y}}^{y}-2 \sigma_{\mathbf{r}}^{z} \sigma_{\mathbf{r}+\mathbf{M}_{z}}^{z}\right) \right].
    \label{eq: HcE2}
\end{align}

\subsection{Real space diagonalization of the Majorana fermion Hamiltonian and  fermionic propagators}\label{sec:RSF}

Using the Kitaev four Majorana fermion representation  of spin \cite{Kitaev2006}, $\sigma_{l}^{\alpha}=i b_{l}^{\alpha} c_{l}$,
 we rewrite the spin Hamiltonian as
\begin{align}
    \mathcal{H}_s=& J_K \sum_{\langle lm \rangle} \frac{i}{2}\eta_{lm} c_{l} c_{m}+\kappa \sum_{\lefta lnm\righta} \frac{i}{2} \eta_{ln} \eta_{mn}c_{l} c_{m} \nn\\
    =& \frac{i}{4} \sum_{\langle lm\rangle} h_{lm } c_{l} c_{m},\label{eq:Hamfermion}
\end{align}
where $\eta_{l m}=i b_{l}^{\alpha} b_{m}^{\alpha}$ is the $Z_2$ link operator on $\alpha$-bond between sites $l$ and $m$ of the honeycomb lattice, and 
$ h_{l m} \equiv  2 J_K \eta_{l m} +2 \kappa\, \eta_{l n} \eta_{m n} $  denote matrix elements of the coupling matrix ${\hat h}$.

The Hilbert space of the fermionic model \refeq{eq:Hamfermion} is larger than that of the spin model; the latter is recovered by imposing the gauge condition, $ b^x_l b^y_l b^z_l c_l |\psi_\tx{s}\rangle=|\psi_{\tx{s}} \rangle$, where $|\psi_\tx{s}\rangle$ is a physical spin state, which is obtained by a projection of a Majorana fermion state $|\Psi_\tx{MF} \rangle$ solved from the Hamiltonian \refeq{eq:Hamfermion} to the physical spin state $|\psi_\tx{s}\rangle$ \cite{Kitaev2006}. The bond operators $\eta_{lm}$ commute with the Hamiltonian \refeq{eq:Hamfermion} and take eigenvalues $\eta_{lm} = \pm 1$. A configuration of $\{ \eta_{lm} \}$ can be understood as $\mathbb{Z}_2$ gauge fields, which corresponds to different gauge choices. The physical spin states, obtained after projecting out the gauge degrees of freedom, is denoted by the eigenvalue of the loop operator $\hat{W_p}$, which in the fermionic representation is expressed as $\hat{W_p} = \prod_{(l,m)\in \tx{edge}(p)} \eta_{lm}$. It can be understood as a gauge invariant Wilson loop operator around a single plaquette $p$. It takes eigenvalue $W_p = \pm 1$. $W_p = - 1$ corresponds to a (gapped) $\pi$-flux excitation on the plaquette in question.  In the following we  will also use notation $ \{\phi_p \}$  to denote a particular flux sector, described by choosing a particular configuration of the bond values $\{{\eta}_{lm}\}$. 



A quantitative description of the finite-temperature  fermionic spectrum is obtained by solving
the free-fermion problem  (\ref{eq:Hamfermion}) in each flux sector exactly.   To this end,  we first specify the link variable  for each bond  and then determine the physically relevant flux sector. After  the flux sector is determined, the Hamiltonian  (\ref{eq:Hamfermion}) can be  solved exactly as a tight-binding model of Majorana fermions \cite{Kitaev2006}.

At $\kappa=0$, the Hamiltonian matrix ${\hat h}$ is anti-symmetric, i.e.\ ${\hat h}^{\top}=-{\hat h}$. Thus, it can be written in a block off-diagonal form 
${\hat h} = 
\left[
    \begin{array}{cc}
                 0 & {\hat M} \\
        -{\hat M}^{\top} &  0
    \end{array}
\right]$. It can be
 block-diagonalized by  the orthogonal matrix ${\hat Q} \in O(2N)$, where $N$ is the number of unit  cells, given by 
 ${\hat Q} = 
\left[
    \begin{array}{cc}
        {\hat u}  & 0 \\
         0 &  {\hat v}
    \end{array}
\right]$, which can be obtained by singular value decomposition. Then,  the orthonormal Majorana modes are given by
\begin{align}
    \gamma_{A, i}&=\sum_{\vrr} u_{\vrr i} c_{A, \vrr}\,,\nonumber\\
    \gamma_{B, i}&=\sum_{\vrr} v_{\vrr i} c_{B, \vrr}\,,\label{transfo1}
\end{align} 
and ${\hat M} = {\hat u}\,{\hat \Omega}\, {\hat v}^{\top}$, where ${\hat \Omega}=\text{diag}(\epsilon_1,\epsilon_2,...,\epsilon_N)$ is the diagonal energy matrix. 
In terms of complex fermions through
\begin{align}
    \beta_{i}=\left(\gamma_{A, i}+i \gamma_{B, i}\right) / 2, \label{eq:transo2}
\end{align}
the Hamiltonian can be written in the canonical form:
 \begin{align}
\mathcal{H}_s = \sum_{i=1}^N\epsilon_{i}\left({\beta}_{i}^{\dagger}{\beta}_{i}-1 / 2\right),\label{eq:HamMF-diag}
\end{align}
where ${\beta}_i$ are complex matter fermions  which label the eigenmodes  with the fermion energies $\epsilon_i$
 for a given flux sector.  From here we can obtain the expressions for the fermion propagators. Since we are interested in finite-temperature  picture, we use  the Matsubara formalism \cite{Mahan}:
 \begin{align}
     &g_i(i\omega_{n})=
     -\langle T_\tau \beta_{{i}}(\tau)\beta_{{i}}^{\dagger}(0)\rangle_{\omega_n}
     = \onefrac{i\omega_{n} - \epsilon_{i}} ,\nonumber \\
    &{\bar g}_i(i\omega_{n} )
    =-\langle T_\tau \beta^{\dagger}_{{i}}(\tau)\beta_{{i}}(0)\rangle_{\omega_n} 
    = \onefrac{i\omega_{n} + \epsilon_{i}},\label{eq:MFgfs}
\end{align}
where $\langle \hat{O}\rangle_{\omega_n}=\int_0^\beta \ud \tau e^{i\omega_n \tau} \langle \hat{O}(\tau) \rangle$, and $T_\tau$ is imaginary time ordering operator. 
  Next, to  describe the diagonalization of the Hamiltonian \refeq{eq:Hamfermion},  we introduce the following matrix form notations. We denote the vector of Majorana fermions as  $\mathbf{C}^{\top}=\left[c_{A,1}, \cdots c_{A,N}, c_{B,1} \cdots c_{B,N}\right]$, and the vector of the Bogoliubov quasiparticles as  $\mathbf{B}^{\dagger}=\left[\beta_{1}^{\dagger},\cdots \beta_{N}^{\dagger}, \beta_{1},\cdots \beta_{N}\right]$. Then the transformations Eq.\ (\ref{transfo1}-\ref{eq:transo2}) can be written as a unitary matrix: $\mathbf{B}^\dagger = \onefrac{\sqrt{2}}\mathbf{C}^\dagger {\hat U}$, where ${\hat U}$ is the unitary matrix that diagonalizes the Hamiltonian matrix $\hat{h}$, and the factor $\onefrac{\sqrt{2}}$ is used to recover the correct anti-commutation relation of the complex fermions.  The vector $\mathbf{B}^\dagger$ has the dimension $2N$ and is naturally divided into the space of creation operators corresponding to the positive energy spectrum and the space of annihilation 
  operators corresponding to the negative energy spectrum.

\subsection{Mixed representation of the  Majorana fermion-phonon (MFPh) coupling}\label{sec:MFPh}
Now  we can  rewrite the coupling Hamiltonian  (\ref{eq: HcE2})  using the Majorana fermion representation of spin as

\begin{align}
    &\mathcal{H}_{\text c}^{E_{2}}=-i \lambda_{E_{2}} \sum_{\mathbf{r}\in A} \left[\phantom{\sqrt{3}}\right. \nn\\
    &2 \sqrt{3}\, \epsilon_{x y}\left(
        \eta_{\vrr,\mathbf{r}+\mathbf{M}_x}c_{\mathbf{r}} c_{\mathbf{r}+\mathbf{M}_{x}}
        \!\!-\!\eta_{\vrr,\mathbf{r}+\mathbf{M}_y}c_{\mathbf{r}} c_{\mathbf{r}+\mathbf{M}_{y}}\right) \!+\! \left(\epsilon_{x x} \!\!-\!\epsilon_{y y}\right) 
    \nn 
    \\
    &\!\!\left(
        \eta_{\vrr,\mathbf{r}+\mathbf{M}_x}c_{\mathbf{r}} c_{\mathbf{r} + \mathbf{M}_{x}}
        \!\!+\!\eta_{\vrr,\mathbf{r}+\mathbf{M}_y}c_{\mathbf{r}} c_{\mathbf{r} + \mathbf{M}_{y}}
        \!\!-\!2 \eta_{\vrr,\mathbf{r}+\mathbf{M}_z}c_{\mathbf{r}} c_{\mathbf{r}+\mathbf{M}_{z}}
    \right) \left.\!\! \right]. \label{eq:HcE2-1}
\end{align}
Next we  will use  a mixed representation, i.e. we  will transform the strain tensor into the momentum space using the long wavelength limit, 
$ \epsilon_{\alpha \beta}(\vrr)=\frac{1}{\sqrt{N}} \sum_{\bf q}\frac{i}{2}\left(q_{\alpha} u_{{\bf q},\beta}+q_{\beta} u_{{\bf q},\alpha}\right)e^{i\vq\cdot\vrr_l}$,
where $\alpha,\beta$ denote $x$ or $y$, but keep Majorana fermions in the real space. The coupling Hamiltonian (\ref{eq:HcE2-1}) then reads 
\begin{align}
     &\mathcal{H}_{\text c} =\onefrac{\sqrt{N}}\sum_{\bf q} V_{\bf q},\nn\\&
     V_{\bf q} =-\frac{i}{2} \sum_{\langle l \in A, m \in B\rangle} c_{l} c_{m}\left(\hat{\lambda}_{{\bf q} , l m}^{\|} \tilde{u}_{q}^{\|}+\hat{\lambda}_{{\bf q} , l m}^{\perp} \tilde{u}_{\bf q} ^{\perp}\right) e^{i {\bf q}  \cdot {\bf r}_l}, \label{eq: Vq} 
    \end{align}
    where the MFPh coupling vertices  are given by
\begin{align}
\hat{\lambda}_{{\bf q} , l m}^{\|}=&
    2i\lambda \left \{
        \delta_{\langle lm\rangle_x}\eta_{l m}\!\left[
         ( c_{\vq} \!+\! \sqrt{3} s_{\vq}) q_{x} \!+\!( \sqrt{3}c_{\vq} \!-\! s_{\vq}) q_{y}
         \right] 
    \right.\nn
\\
 +&  \delta_{\langle lm\rangle_y}\eta_{l m}\left[
    ( c_{\vq}- \sqrt{3} s_{\vq}) q_{x}+(- \sqrt{3} c_{\vq}+ s_{\vq}) q_y
\right] \nonumber
\\
+&\left.  \delta_{\langle lm\rangle_z}\eta_{l m}\, 2 \left(
    - c_{\vq} q_{x}+ s_{\vq} q_{y}\right) \right\},\\
\hat{\lambda}_{{\bf q} , l m}^{\perp}=&
    2i\lambda \left \{ 
        \delta_{\langle lm\rangle_x}\eta_{l m} \!\left[
            ( \sqrt{3}c_{\vq} \!- \! s_{\vq})q_x \!+\! (\!-\! c_{\vq}\!-\! \sqrt{3} s_{\vq}) q_y
        \right] 
    \right.\nonumber\\
 +&  \delta_{\langle lm\rangle_y}\eta_{l m}\left[
    (- \sqrt{3} c_{\vq}- s_{\vq})q_{x}+(-c_{\vq}+ \sqrt{3} s_{\vq})q_{y}
\right] 
\nonumber \\
 + &\left.\delta_{\langle lm\rangle_z}\eta_{l m}\,2 \left(
    s_{\vq}q_x +   c_{\vq} q_y\right) \right\}. 
\end{align}
Here, for the compactness of the equations we denote $c_{\vq}= \cos \theta_{\vq}$ and $s_{\vq}= \sin \theta_{\vq}$,
and $\delta_{{\langle lm\rangle}_\alpha}=1$ when $\langle lm\rangle$ is the nearest neighboring link of type
$\alpha\in\{x, y,z\}$, and zero otherwise. Since $l$ and $m$ always belong to different  A and B sublattices, we can write
the coupling Hamiltonian \refeq{eq: Vq}  as
\begin{align}
    V_{\bf q} = -\frac{i}{4} \sum_{ l, m}
    c_l c_m \Lambda^{\mu}_{{\bf q} , lm} \tilde{u}_{\bf q} ^{\mu} \label{eq: Vq2}
\end{align}
where the coupling matrices $\Lambda^\mu_{{\bf q}, lm},\ \mu=\|, \perp$,  in the sublattice matrix representation  are given by
\begin{align}
    &\Lambda_{{\bf q}, lm}^{\mu} =   \label{eq: sublattice_block}\\ 
    &\left[\begin{array}{cccccc}        
    &   &   &     & {\phantom\cdots} &   \\
    & O &   &     & {\hat{\lambda}}_{{\bf q}}^{\mu }     &      \\
    &   &  &    & {\phantom \cdots} &           \\
    & {\phantom \vdots} &{\phantom \vdots}      	&      & &           \\
    & {\hat{\lambda}}_{{\bf q}}^{\mu \dagger} &{\phantom \vdots}  &  & O & \\
    & {\phantom \vdots} &{\phantom \vdots}                                     &  	   & &\\
    \end{array}\right]_{lm}
    \!\!\!\!\!\bigodot
    \left[\begin{array}{cccccc}
         & & & \cdots & \cdots &\cdots\\
         & O& & \cdots & e^{i{\bf q}\cdot {\bf r}_l} &\cdots \\
         & & & \cdots & \cdots &\cdots \\
        \vdots & \vdots &\vdots                         & & &\\
        \vdots & e^{i{\bf q}\cdot {\bf r}_m} &\vdots   & & O&\\
        \vdots & \vdots &\vdots                         & & &\\
    \end{array}\right]_{lm},\nn
\end{align}
where 
 $\bigodot$ is element-wise multiplication.
Notice that in the spin-lattice coupling \refeq{eq: Vq}, the plane wave phase factor is applied only on the $A$ sites. To make it invariant under the $C_{6v}$, a further symmetrization of  the coupling matrix between  between $A$ and $B$ sublattices is required.  By performing this symmetrization,
 we obtained the symmetrized  spin-lattice coupling between the eigenmodes of the Hamiltonian:
\begin{align}
    V_{ \bf q}&=-\frac{i}{2} \sum_{\langle l \in A, m \in B\rangle} c_{l} c_{m}\hat{\lambda}_{{\bf q}, l m}^{\mu} \tilde{u}_{\bf q}^{\mu}\,\oh (e^{i {\bf q} \cdot {\bf r}_l} + e^{i{\bf q} \cdot {\bf r}_m}) \nn
  \\
    &= -\frac{i}{4}  \sum_{ l, m }  
    c_l c_m {\Lambda}^{s,\mu}_{{\bf q}, lm} \tilde{u}_{\bf q}^{\mu}
    = -\frac{i}{2} {\bf B}^\dagger \tilde{{\Lambda}}^\mu_{\bf q} {\bf B}\, \tilde{u}_{\bf q}^\mu \label{eq: Vq3}
\end{align}
where ${\tilde\Lambda}_{\bf q}^\mu \equiv {\hat U}^\dagger \Lambda^{s,\mu}_{\bf q} {\hat U}$ is the symmetrized coupling matrix, whose entries are the coupling vertices between two fermion eigenmodes.   The coupling matrices $\tilde{\Lambda}_{\bf q}^\mu$  can also be divided into four blocks according to the division into the creation and annihilation subspace:
\begin{align}
    \tilde{\Lambda}_{\bf q}^\mu & \equiv\left[\begin{array}{ll}
    \tilde{\Lambda}_{{\bf q},11}^\mu & \tilde{\Lambda}_{{\bf q},12}^\mu \\
    \tilde{\Lambda}_{{\bf q},21}^\mu & \tilde{\Lambda}_{{\bf q},22}^\mu
    \end{array}\right]. \label{eq: sub_couplingM}
\end{align}
These coupling matrices will be used in the final expression of the polarization bubble, which we will derive in the next section.

\section{Phonon polarization bubble } \label{sec:III}
Next we  discuss  the effect of the spin-lattice coupling on the  phonon dynamics.
By calculating the one-loop phonon self-energy $\Pi^{\mu\nu}_{ph}(\vq,\Omega)$, the renormalization to the sound velocity, mixing of the transverse and longitudinal phonon modes, and attenuation/absorption of sound wave may be obtained.
 Using \refeq{eq: Vq3}, the phonon self-energy or, equivalently, the polarization bubble, can be expressed as
\begin{align}
    \Pi^{\mu\nu}({\bf q}, \tau) = 
    \left\langle T_{\tau}\left({\bf B}^{\dagger} \tilde{\Lambda}_{\bf q}^{\mu} {\bf B}\right)\left(\tau\right)\left({\bf B}^{\dagger} \tilde{\Lambda}_{-{\bf q}}^{\nu} {\bf B} \right)\left(0\right)\right\rangle, \label{eq: Pi_t}
\end{align}
where the factor $\onefrac{2}$  from the \refeq{eq: Vq3} has been  absorbed into the definition of $\tilde{\Lambda}_{\bf q}$. By using the Wick's theorem, the polarization bubble can be explicitly written as:
\begin{align}
   \!\! &\Pi^{\mu \nu}\!\left({\bf q}, \tau \right) \!=\! \left\langle 
        T_\tau{{\bf B}}_{k}^{\dagger} \!\left(\tau \right){ {\bf B}}_{m}\!\left(0 \right)
    \right\rangle \!
    \left\langle 
        T_\tau{ {\bf B}}_{l} \!\left(\tau \right){ {\bf B}}_{n}^{\dagger}\!\left(0 \right)
    \right\rangle \!
    \tilde{\Lambda}_{k l, { \vq}}^{\mu} \tilde{\Lambda}_{n m, \!-\!{\bf q}}^{\nu} \nn
    \\
    &\quad -\left\langle
        T_\tau{{\bf B}}_{k}^{\dagger}\!\left(\tau \right) { {\bf B}}_{n}^{\dagger}\!\left(0 \right)
    \right\rangle
    \left\langle
        T_\tau{{\bf B}}_{l}^{\phantom{\dagger}}\!\left(\tau \right) { {\bf B}}_{m}\!\left(0 \right)
    \right\rangle
    \tilde{\Lambda}_{k l, {\bf q}}^{\mu} \tilde{\Lambda}_{m n, -{\bf q}}^{\nu\,\top}, \label{eq:BBBB}
\end{align}
where  the indices $k,l,m,n$ range from $1$ to $2N$.
Performing the Fourier transform and using the \refeq{eq: sub_couplingM},  we then get
\begin{align}
   & \Pi^{\mu\nu} ({\bf q}, i\Omega_m) 
    = \Tr  \left[ \right. \nn
    \\&
    \! \bar{g}\!\left(i \omega_{n_1}\!\right) \tilde{\Lambda}^{\mu}_{{\bf q},11} \ g\!\left(i \omega_{n_{2}}\!\right) \tilde{\Lambda}^{\nu}_{{\bf q},11} 
    \!+\! g\!\left(i \omega_{n_1}\!\right) \tilde{\Lambda}^{\mu}_{{\bf q},21}\ g\!\left(i \omega_{n_{2}}\!\right) \tilde{\Lambda}^{\nu}_{{\bf q},12}
    \nonumber
    \\
    \!+& \bar{g}\!\left(i \omega_{n_{1}}\!\right) \tilde{\Lambda}^{\mu}_{{\bf q},12} \ \bar{g}\!\left(i \omega_{n_{2}}\!\right) \tilde{\Lambda}^{\nu}_{{\bf q},21}
    \!+\!g\!\left(i \omega_{n_{1}}\!\right) \tilde{\Lambda}^{\mu}_{{\bf q},22}\ \bar{g}\!\left(i \omega_{n_{2}}\!\right) \tilde{\Lambda}^{\nu}_{{\bf q},22} \nonumber
    \\ 
    \!-& \bar{g}\!\left(i \omega_{n_{1}}\!\right) \tilde{\Lambda}^{\mu}_{{\bf q},11}\ g\!\left(i \omega_{n_{2}}\!\right) \tilde{\Lambda}^{\nu \top}_{{\bf q},22}
    \!-\!g\!\left(i \omega_{n_{1}}\!\right) \tilde{\Lambda}^{\mu}_{{\bf q},21}\ g\!\left(i \omega_{n_{2}}\!\right) \tilde{\Lambda}^{\nu \top}_{{\bf q},12} \nonumber
    \\  
    \!-& \left.\!\!\bar{g}\!\left(i \omega_{n_{1}}\!\right) \tilde{\Lambda}^{\mu}_{{\bf q},12}\ \bar{g}\!\left(i \omega_{n_{2}}\!\right) \tilde{\Lambda}^{\nu \top}_{{\bf q},21}
    \!-\!g\!\left(i \omega_{n_{1}}\!\right) \tilde{\Lambda}^{\mu}_{{\bf q},22}\ \bar{g}\!\left(i \omega_{n_{2}}\!\right) \tilde{\Lambda}^{\nu \top}_{{\bf q},11}\right], \label{eq: Pi_expression}
\end{align}
where $\Tr[...]$ now sums over the Matsubara frequencies $i\omega_n$ as $T\sum_n$, and 
the energy conservation constraint $\Omega_{m}-\omega_{n_{1}} -\omega_{n_{2}}=0$ is imposed.  Here, $g(i\omega_n) = \tx{diag}(\cdots  g_i(i\omega_n) \cdots)$ and $\bar{g}(i\omega_n) =  \tx{diag}(\cdots  \bar{g}_i(i\omega_n) \cdots)$ are the diagonal matrices. 
  $ \Pi^{\mu\nu} ({\bf q}, i\Omega_m) $ can   also be  conveniently  written  in a matrix form:
\begin{align}
     & \Pi^{\mu\nu}({\bf q}, i\Omega_m) =-\Tr\left[ \right. \label{eq:PI} \\
     & \left.\!\! G_{1}(i\omega_{n_1}) \tilde{\Lambda}_{\bf q}^{\mu}G_{1}^{*}(i\omega_{n_2}) \tilde{\Lambda}_{-{\bf q}}^{\nu}
     +G_{2} (i\omega_{n_1}) \tilde{\Lambda}_{\bf q}^{\mu}G_{2}(i\omega_{n_2})\tilde{\Lambda}_{-{\bf q}}^{\nu \top}\right],\nn 
\end{align}
where 
\begin{align}
    G_1 (i\omega_{n})\equiv 
    \left[\begin{array}{ll}
        \overline{g}(i\omega_{n})& O \\
        O & g(i\omega_{n})
    \end{array}\right], \nonumber \\
G_2 (i\omega_{n})\equiv 
    \left[\begin{array}{ll}
        O & g (i\omega_{n})\\
        \overline{g} (i\omega_{n})& O
    \end{array}\right], \label{eq:29}
\end{align}
and  $\overline{g}(i\omega_n) = -g^*(i\omega_n)$.
From  \refeq{eq: Pi_expression}, it is clear that the frequency dependence only appears in the denominator of the fermion propagators.  Thus, we  can  explicitly   sum  over the Matsubara frequences,  then
take analytical continuation: $i \Omega_{m} \rightarrow \Omega+i \delta$ and obtain the final expression of phonon polarization bubble:
\begin{align} 
    &\Pi^{\mu \nu}({\bf q}, \Omega) =  \onefrac{ N} 
    \sum_{ij} \label{eq:polarization}
    \\
    & \left[ P_{i j}^{\bar{g} g} 
    \left[\tilde{\Lambda}^{\mu}_{{\bf q},11}\right]_{ i j}
     \left[\tilde{\Lambda}^{\nu}_{{\bf q},11}\right]_{ i j}
    +P_{i j}^{gg}
    \left[\tilde{\Lambda}^{\mu}_{{\bf q},21}\right]_{ i j}
     \left[\tilde{\Lambda}^{\nu}_{{\bf q},12}\right]_{ i j} \right. \nn
    \\ & 
    + P_{i j}^{\bar{g} \bar{g}}
     \left[\tilde{\Lambda}^{\mu}_{{\bf q},12}\right]_{ i j}
     \left[\tilde{\Lambda}^{\nu}_{{\bf q},21}\right]_{ i j}
    +P_{i j}^{g \bar{g}} 
    \left[\tilde{\Lambda}^{\mu}_{{\bf q},22}\right]_{ i j}
    \left[\tilde{\Lambda}^{\mu}_{{\bf q},22}\right]_{ i j}
     \nn
    \\ &  
    -P^{\bar{g}g}_{i j} 
    \left[\tilde{\Lambda}^{\mu}_{{\bf q},11}\right]_{ i j}
    \left[\tilde{\Lambda}^{\nu}_{{\bf q},22}\right]_{ i j}
     -P_{i j}^{g g} \left[\tilde{\Lambda}^{\mu}_{{\bf q},21}\right]_{ i j}
     \left[\tilde{\Lambda}^{\nu}_{{\bf q},12}\right]_{ i j} \nn
    \\
    & \left.\!\! 
    -P^{\bar{g} \bar{g}}_{i j} \left[\tilde{\Lambda}^{\mu}_{{\bf q},12}\right]_{ i j}\left[\tilde{\Lambda}^{\nu}_{{\bf q},21}\right]_{ i j}
    -P_{i j}^{g \bar{g}}\left[\tilde{\Lambda}^{\mu}_{{\bf q},22}\right]_{ i j}\left[\tilde{\Lambda}^{\nu}_{{\bf q},11}\right]_{ i j} 
    \right],  \nn
\end{align}
where $P^{g\overline{g}}_{ij},  P^{\bar{g} g}_{ij},   P^{gg}_{ij}$ and $P^{\bar{g}\bar{g}}_{ij}$ are obtained from the Matsubara summation over the frequencies  and are  explicitly given   in Eq. (\ref{eq:PPPP}) of App.~\ref{app: eval_bubble}. We will see in the  next section that  
since $P^{g\overline{g}}_{ij},  P^{\bar{g} g}_{ij},   P^{gg}_{ij}$ and $P^{\bar{g}\bar{g}}_{ij}$ encode the information of the fermionic energy spectrum, they  determine the intensity of the Majorana fermion-phonon scattering and its dependence on temperature. We will also see that
 the matrix elements $\tilde{\Lambda}^\mu_{{\bf q},ij}$, describing the coupling between the $i$-th and $j$-th fermionic eigenmodes and the acoustic phonon with wavevector ${\bf q}$ and the polarization $\mu$, are responsible
for the angular  dependence of the  Majorana fermion-phonon scattering.

\begin{figure}
	\centering
	\includegraphics[width=1.0\columnwidth]{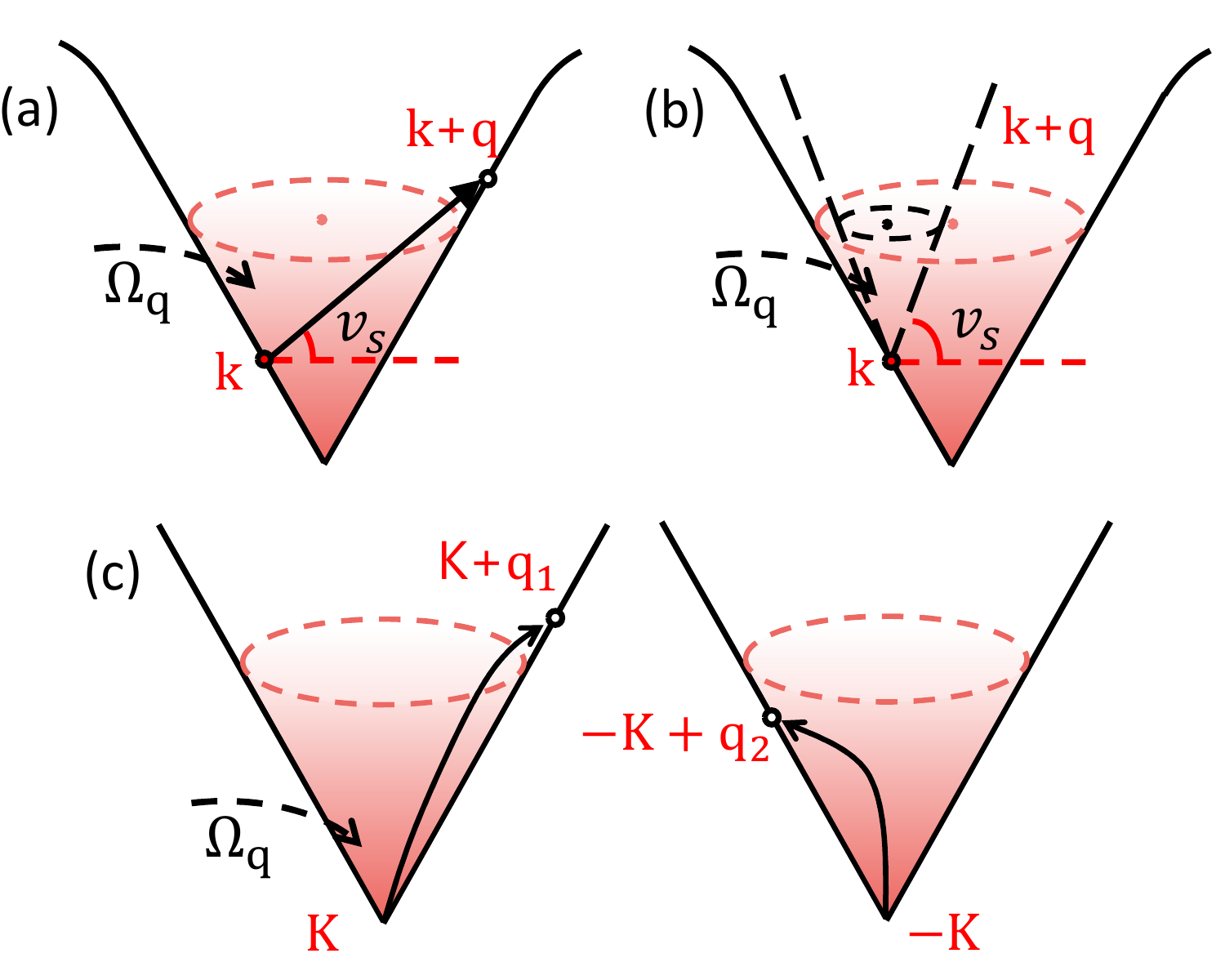}
     \caption{The illustration of different scattering channels near the bottom of the Dirac cone with different acoustic phonon velocities $v_s$: (a) ph-channel with $v_s < v_F$, (b) ph-channel with $v_s > v_F$, (c) pp-channel with $v_s>v_F$. Here, $v_F$ is the Fermi velocity visualized by the slope of the Dirac cone. The phonon energy is $\Omega_\vq = v_s|\vq|$. In scenario (a), the kinematic constraint is satisfied for a hopping from a fermionic mode ${\bf k}$ to ${\bf k}+{\bf q}$. In scenario (b), the dashed cone
   illustrating  a  hopping from a fermionic mode $\vk$ to any of the possible final states that satisfies the kinematic constraint
   has no intersection with the fermionic  cone (solid line) since $v_s > v_F$. Thus, this scenario is not possible.  }	\label{cone2}
\end{figure}

\section{The sound attenuation coefficient}\label{sec:IV}
The quantitative description of the  attenuation process can be obtained  
through the lossy acoustic wave function  which  
decays with distance away from the driving source as
\begin{align}
    \mathbf{u}(\mathbf{r}, t)=\mathbf{u}_{0} e^{-\alpha_{s}(\mathbf{q}) r} e^{i(\Omega t-\mathbf{q} \cdot \mathbf{r})},
\end{align}
where $\alpha_s(\vq)$ is the attenuation coefficient defined as the inverse of the phonon mean free path, which can be computed
from the imaginary part of the diagonal components of the phonon self-energy \cite{Plee2011,Ye2020}:
\begin{align}
    \alpha^{\mu}_{s}(\mathbf{q}) \propto-\frac{1}{v_{s}^{2} q} \operatorname{Im}\left[\Pi_{\mathrm{ph}}^{\mu\mu}(\mathbf{q}, \Omega)\right]_{\Omega=v_{s} q}, \label{alpha-anal}
\end{align}
where $\mu$ is the polarization component. In  Ref.~\cite{Ye2020} we have  shown
that at low temperatures below the energy scale of the flux gap,  the sound attenuation is determined by the decay of a phonon into a pair of Majorana fermions, with the attenuation rate linear in temperature due to the  density of states that scales linearly in energy at the Dirac points. [Note that this  only applies when $v_s<v_F$.]
The question on how  the presence of thermally excited fluxes  modifies this picture is  addressed  in this section.

\begin{figure*}
        \centering
        \includegraphics[width=\textwidth]{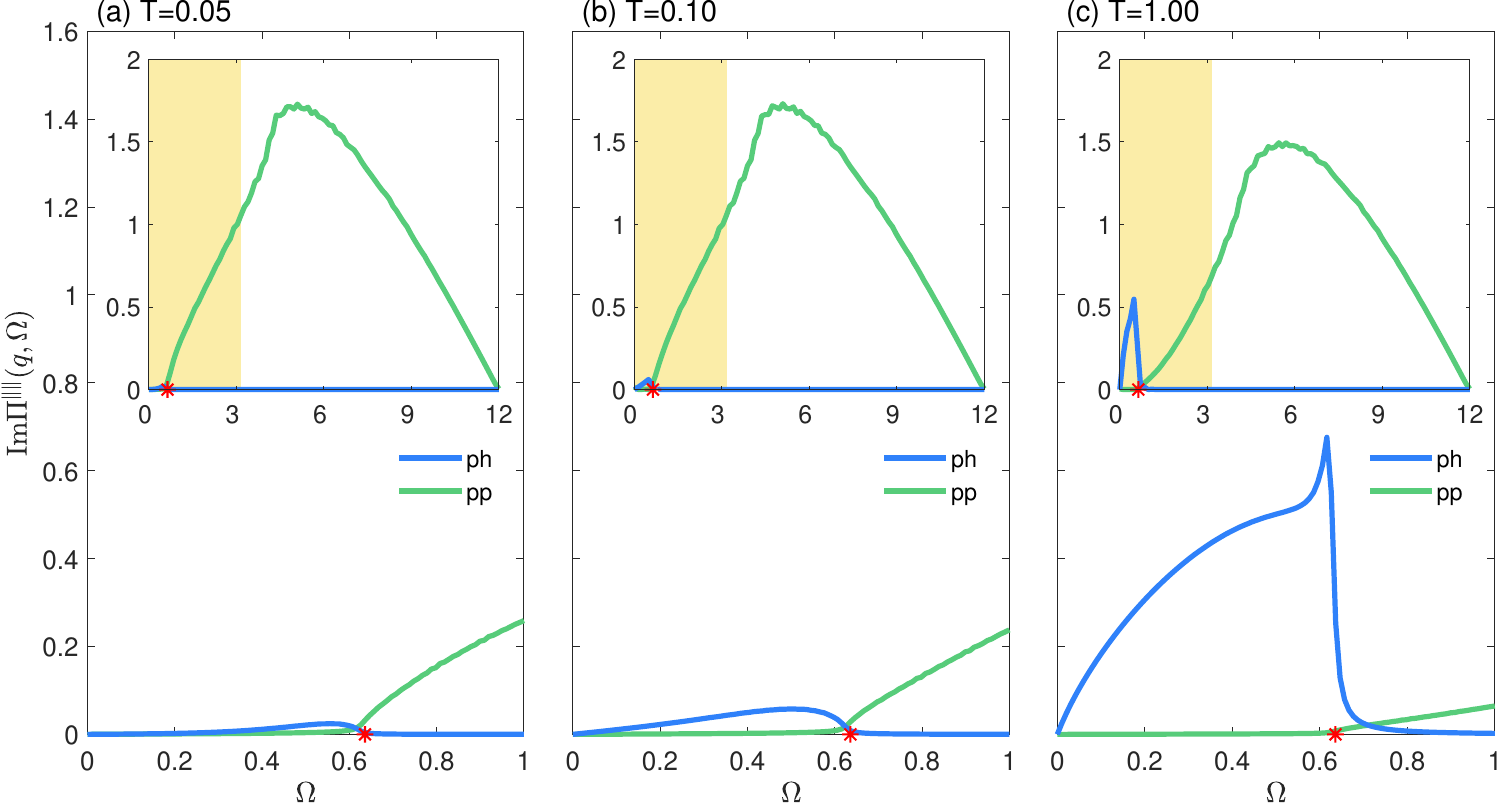}
        \caption{The imaginary part of the phonon self-energy, $\text{Im}\Pi^{\|\|}(\vq, \Omega )$   as a function of the phonon  frequency $\Omega$  and  
        fixed phonon momentum ${\bf q}=(0.1 \pi, 0)$
        computed assuming the zero-flux sector for various temperatures. The ph-channel and pp-channel are shown as blue and green curves, respectively. The calculations are performed  on the finite size  lattice with $N_1 = N_2 = 1000$. 
        We use $\delta= 0.01$ for  imaginary energy broadening. The red star at the frequency $\Omega^*=\epsilon_{{\bf K} + \vq}=v_F|{\bf q}|$, where ${\bf K}$ denoting one of the Dirac points, corresponds to the upper bound for the ph-continuum and the lower bound of the pp-continuum. The inset in each of the panel  shows  the ph- and pp-contributions to $\text{Im}\Pi^{\|\|}(\vq, \Omega )$ in a wide range of the phonon frequencies. The shaded region in each of the  insets highlights the range of the phonon frequencies in which the long wavelength  description of the phonon media is applicable.  The frequency, $\Omega$, and temperature, $T$, are measured
  in units of the Kitaev interaction, $J_K$.}
        \label{intensity}
\end{figure*}

\begin{figure*}[htbp]
        \centering
        \includegraphics[width=0.95\textwidth]{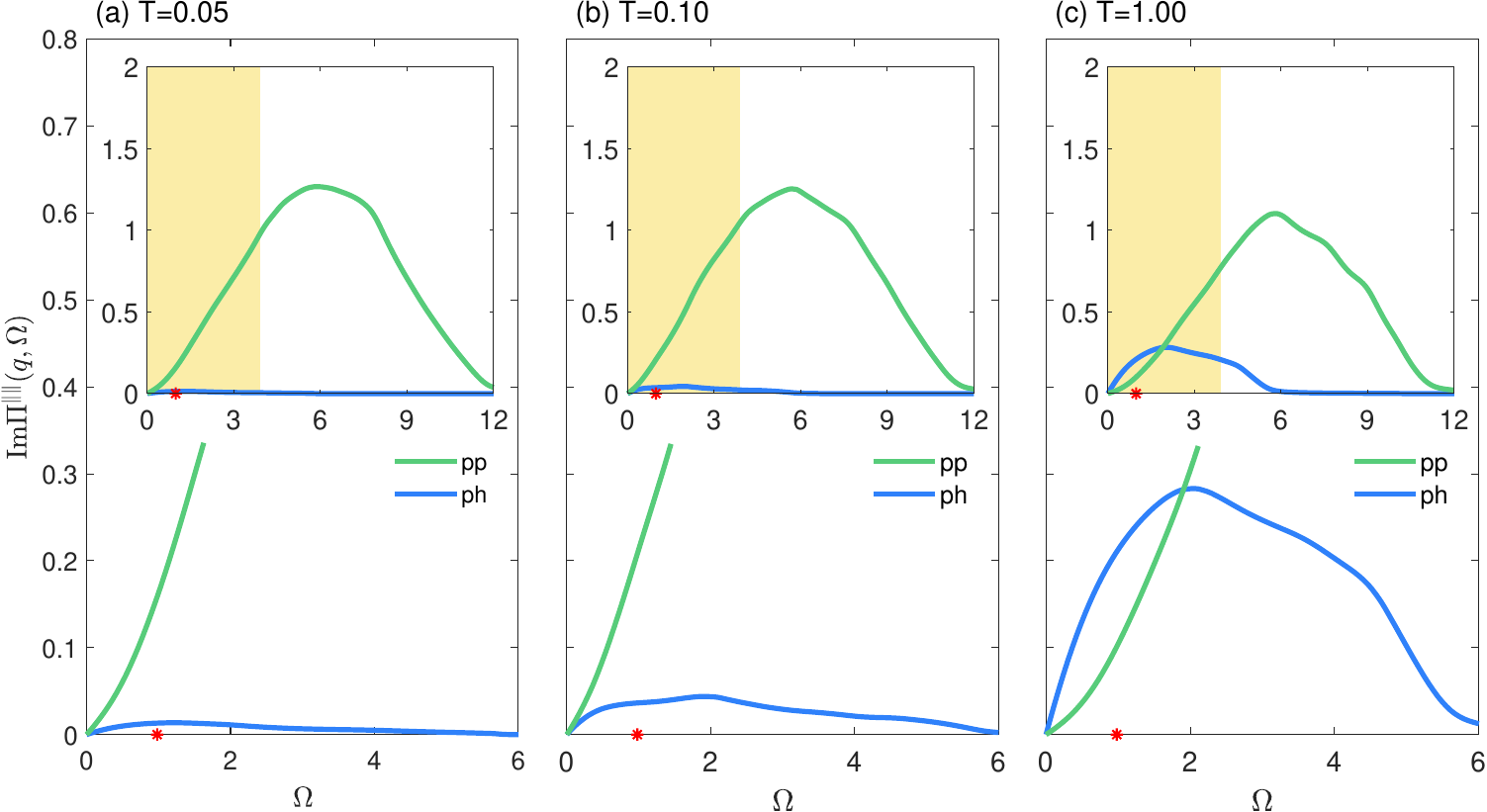}
        \caption{The imaginary part of the phonon self-energy $\text{Im}\Pi^{\|\|}(\vq, \Omega )$   as a function of the phonon  frequency $\Omega$  and  
        fixed phonon momentum ${\bf q}=(0.1 \pi, 0)$
        computed  using \refeq{eq:polarization} for (a) $T=0.05$, (b) $T=0.1$ and (c) $T=1.0$.  For every temperature, 200 inhomogeneous flux configurations  are sampled with the help of the  strMC algorithm (see \refapp{app: sMC}) implemented on the finite size lattice with $N_1 = N_2 = 32$.
         The ph-channel and pp-channel are shown as blue and green curves, respectively.   We use $\delta= 0.2$ for the imaginary energy broadening. The red star at the frequency $\Omega^*=\epsilon_{{\bf K} + \vq}=v_F|{\bf q}|$, where ${\bf K}$ denoting one of the Dirac points, corresponds to the upper bound for the ph-continuum and the lower bound of the pp-continuum in the zero-flux sector. The inset in each of the panel  shows  the ph- and pp-contributions to $\text{Im}\Pi^{\|\|}(\vq, \Omega )$ in a wide range of the phonon frequencies. The shaded region in each of the  insets highlights the range of the phonon frequencies in which the long wavelength  description of the phonon media is applicable. The frequency, $\Omega$, and temperature, $T$, are measured
  in units of the Kitaev interaction, $J_K$. }
        \label{intensity_MC}
\end{figure*}

\subsection{The kinematic constraints in phonon dynamics}
\label{subsec: basic_picture}

Existence of the  coupling between the Majorana fermions and phonons is necessary but  not  a sufficient condition for the 
 decay of the phonon into  the low-energy Majorana  fermions modes.   In the case of the translationally invariant system, the decay rate is  governed by the kinematic   conditions reflecting the conservation of energy and  momentum.    At low temperatures, when the Kitaev spin liquid  can be considered in its ground state zero-flux sector,  both energy and momenta constraints are present. 
  At finite temperatures,
  the  thermally excited  fluxes destroy the translational symmetry such that a fermionic momentum is not any more a good quantum number. Thus,  at finite temperatures the decay rate is determined by a weaker kinematic conditions. In addition, the disorder from the thermal fluxes destroys the  Dirac cones of the Majorana fermions and  flattens their  density of states over the whole energy range, so that the scattering of the phonons is happening on a very different manifold of  low-energy fermionic states.
 

In this section, we analyze the phonon decay in the zero-flux sector, which dominates at low temperatures.  
The relative strength of the sound velocity, $v_s$, and the Fermi velocity, $v_F$, that characterizes the slope of the low-energy  Majorana Dirac cone defines  the phase space for the decay and determines  whether the decay happens in  the particle-hole (ph-) or in the particle-particle (pp-) channel.  Here, by particle and hole, we mean if the state $i$ with energy $\epsilon_i>0$ in Eq.~\eqref{eq:HamMF-diag} is occupied or empty. In other words, the particle number refers to that of the complex fermion $\beta_i$.  
   
The  kinematic constraint for a ph-process, in which a phonon mode with $\Omega_\vq = v_s|\vq|$ scatters a   fermion from  the state  at ${\bf k}$ to  the state at ${\bf k}+{\bf q} $  is shown schematically in \reffg{cone2} (a). As the maximum energy difference between the two states is $v_F|\vq|$, the ph-process is allowed kinematically only when $v_s<v_F$. Moreover, since in order for this process to happen,  some positive energy state, e.g.\ $\beta_{\kv}^{\dagger}|0\rangle$  shown in \reffg{cone2} (a), must be occupied, finite temperature is required.

For a  pp-process, illustrated in \reffg{cone2} (c), in which a phonon decays into two  fermions with positive energies, the kinematic constraints require: 
\begin{align}
    \Omega_\vq  & = \epsilon_{{\bf K}+\vq_1} + \epsilon_{-{\bf K}+\vq_2}  = v_F (|\vq_1| + |\vq_2|)  \nn\\
    &\geq v_F |\vq_1 + \vq_2| = v_F|\vq|, \label{eq: pp_constraint}
\end{align}
where ${\bf K}+\vq_1$, $-{\bf K}+\vq_2$ are, respectively,  the momenta of the pair of fermion particles produced by the phonon, and we have  expanded the energy
of a single fermion  $\epsilon_{{\bf K}+\vq} = 2J_K |\sum_{\alpha =
x,y,z}  e^{i ({\bf K}+\vq) \cdot {\mathbf{M}}_{\alpha}}|\simeq v_F |\vq|$ near the corresponding Dirac points. 
Thus, the pp-process is lower bounded by $v_F|\vq| = \epsilon_{K+\vq}$.  Clearly, when
 $v_s < v_F$ the phonon doesn't have enough energy to produce a pair  of fermion particles, and,  resultantly, in this limit the  low-temperatures sound attenuation is mainly caused by the ph-scattering.  On contrary,  when $v_s \geqslant v_F$, the constraint \refeq{eq: pp_constraint} is easily satisfied, leading to the  decay of the phonon into a pair  of Majorana fermions. Unlike the ph-process, the pp-process does not require  a finite occupation of the fermionic states and, in principle, can happen at  zero temperature (when phonons are externally pumped into the system) as long as the incident phonon has enough energy to excite a pair of particles. 
   As illustrated in \reffg{cone2} (b),  when $v_s \geqslant v_F$ the  kinematic constraints for the ph-process can not be satisfied,  an thus the sound attenuation coefficient is entirely determined by  the pp-scattering.


\reffg{intensity} presents the imaginary part of the longitudinal component of the phonon self-energy, $\text{Im}\Pi^{\|\|} (\vq,\Omega)$, as a function of phonon  frequency  in the  zero-flux sector   for (a) $T=0.05$, (b) $T=0.1$ and (c) $T=1.0$.   The  contributions from the ph-channel and pp-channel  computed at fixed phonon momentum ${\bf q}=(0.1 \pi, 0)$ are  plotted with  blue and green curves, respectively. 

The main  panels of  \reffg{intensity} show  the ph- and pp-contributions to $\text{Im}\Pi^{\|\|} (\vq,\Omega)$  computed in the low-frequency region,  while  the inset in each of the panel  shows  them in a wider range of the phonon frequencies. The red star at the frequency $\Omega^*=v_F |{\bf q}|=0.1\pi v_F$ corresponds to the upper bound for the ph- and the lower bound of the pp-continuum of scattering.  
 By comparing  the panels  (a), (b) and (c) of  \reffg{intensity}, we can   see that the  magnitude of the pp-channel  contribution into $\text{Im}\Pi^{\|\|}(\vq, \Omega )$ is almost  independent on the  temperature. This is because the pp-channel does not require a  finite particle population and  the  decay of the phonon into a pair of Majorana fermions can happen even at zero temperature, as long as the incident phonon has enough energy to excite a pair of particles. On the other hand, the ph-channel requires a finite population of the fermionic states, and its intensity increases with increasing temperature  leading to larger  population of the low-energy with fermionic states.  Note, however,
 that with further increase of temperature, there will be a critical temperature at which the particle and the hole population become balanced, and thus above this  temperature  the scattering probabilities will be suppressed. Clearly, at  $T\to \infty$, the intensities of both the ph- and pp-channels will decay to zero due to the Pauli exclusion principle.  Similar plots for the zero-flux sector and the random flux sectors were shown in Ref.\ \cite{Metavitsiadis2020}.


 \subsection{Phonon dynamics in the presence of thermal fluxes} \label{subsec:IVB}
  
In the inhomogeneous  thermal flux sectors, which are relevant at elevated temperatures, the constraint of momentum conservation is relaxed. Therefore,  at finite temperatures when flux proliferates, the decay of a phonon involves both  the ph- and pp-processes.  
In order to illustrate this, in \reffg{intensity_MC} we plot
the imaginary part of the phonon self-energy computed with \refeq{eq:polarization}  as a function of phonon energy in the thermal flux sectors  for (a) $T=0.05$, (b) $T=0.1$ and (c) $T=1.0$.  Again we focus on the diagonal longitudinal 
component of the polarization bubble \ $\text{Im}\Pi^{\|\|} (\vq,\Omega)$.  The sampling over flux configurations is performed with a help of the stratified Monte Carlo (strMC) 
algorithm (see \refapp{app: sMC} for details) which exploits
 the fact that the energy of each flux configuration  can be computed exactly by diagonalizing the quadratic Majorana Hamiltonian (\ref{eq:HamMF-diag}). The results are computed at fixed phonon momentum ${\bf q}=(0.1 \pi, 0)$ and  are averaged over 200 inhomogeneous flux configurations  for each data point.
In all panels, the red star at the phonon frequency $\Omega^*=v_F |{\bf q}|=0.1\pi v_F$ marks the critical value  of the phonon frequency below which in the zero-flux sector the phonon decays into the ph-continuum  and above which into the pp-continuum.
 
 \reffg{intensity_MC} shows the effects of  the relaxation of the kinematic constraints  and the modifications of the fermionic spectrum   due to the thermal flux. In particular, we can clearly see that 
  the contribution from  the phonon decay into the ph-continuum significantly increases with temperature, while the contribution from the decay  into the  pp-continuum remains almost unchanged. 

 \reffg{intensity_MC} (a) shows $\text{Im}\Pi^{\|\|}(\vq, \Omega )$  computed at temperature $T=0.05$,  which is well  below the flux gap.  At this temperature, the MC  sampling is done predominantly in the zero-flux sector or in a very low-density flux sectors.  Thus,  the pp- and the ph-contribution to $\text{Im}\Pi^{\|\|}(\vq, \Omega )$  should be almost identical  to those shown in \reffg{intensity} (a). Nevertheless, we can see in   \reffg{intensity_MC} (a)  that  the contribution from the pp-channel  at frequencies below $\Omega^*$ is not zero and is notably larger than the contribution from the ph-channel, which violates the kinematic constraints discussed above. This unphysical nonzero pp-channel can be attributed to a finite-size effect, since here we consider much smaller system with $N_1 = N_2 = 32$ in order to perform MC calculation, while in \reffg{intensity} we have used the system  with $N_1 = N_2 = 1000$.
 At higher temperatures shown in \reffg{intensity_MC}  (b) and (c), the flux proliferates, and the ph-continuum is smeared and further extends into the higher frequency region  beyond $\Omega^*$ due to the relaxation of the kinematic constraints. For the same reason, the  pp-continuum could also extend to the lower frequency region below $\Omega^*$. Notice, however,  large finite-size effects in this region. 
Therefore, in the region $\Omega < \Omega^*$, 
the ph-channel of the scattering  is  the dominant over pp-channel.  Thus, when $v_s < v_F$   we will consider only the  contribution from  the ph-channel.  When $v_s \geqslant v_F$, which corresponds to $\Omega \geqslant\Omega^*$ (but not too large that the long-wavelength approximation remains valid), both  the  ph- and the pp-channels contribute similarly into the phonon scattering, so  we will account for both contributions.

 
 \begin{figure}
	\centering
	\includegraphics[width=1.0\columnwidth]{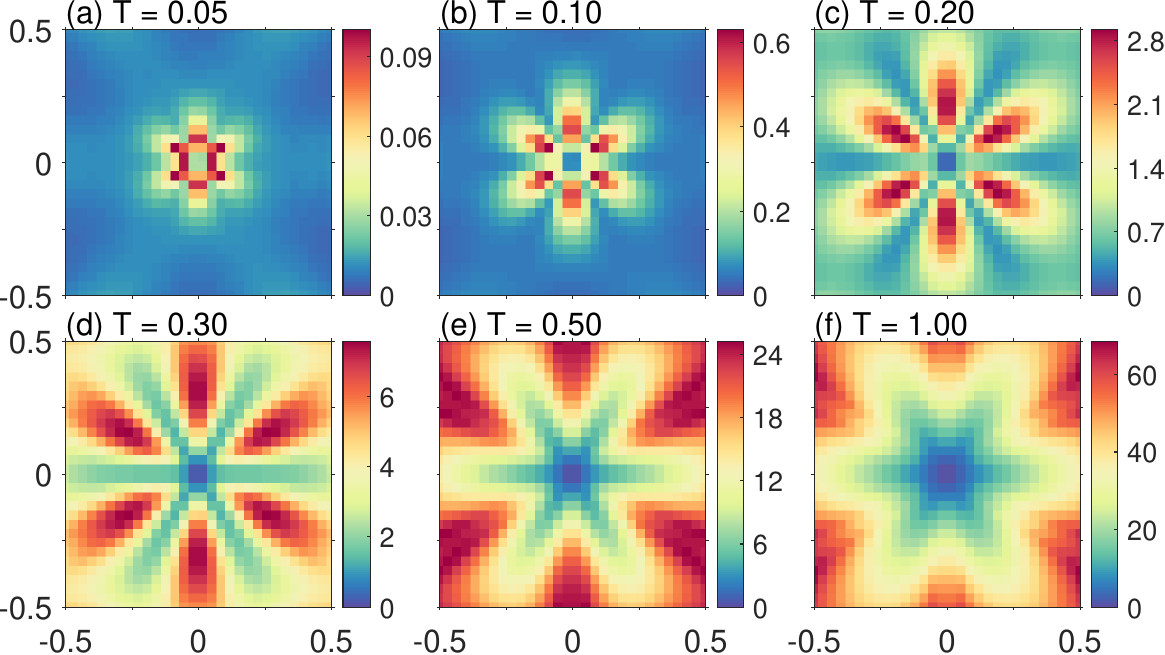}
     \caption{Case $v_s<v_F$: The temperature evolution of the sound attenuation coefficient  $ \alpha_s^{||} ({\bf q})$  computed in the zero-flux sector.  The diagonalization of the Majorana fermion Hamiltonian \refeq{eq:Hamfermion} is  done in the momentum space on the lattice with  $N_1= N_2 = 500$. The phonon momentum ${\bf q}$ belongs to the region $(q_x, q_y) \in [-0.5\pi, 0.5\pi]^2 $.  In the calculations, we set $v_s =0.1 v_F$ and the imaginary energy broadening  $\delta = 0.2$.Temperature is measured in units of the Kitaev interaction, $J_K$.}	\label{kspace_attenuation_large}
\end{figure}

\subsection{Numerical results for the sound attenuation coefficient} \label{subsec:IVC}

In this section, we will  present numerical results for the temperature evolution of the sound attenuation coefficient given by Eq.(\ref{alpha-anal}).  We will consider two cases:  $v_s < v_F$   and  $v_s \geqslant v_F$  corresponding, respectively, to $\Omega<\Omega^*$ and $\Omega>\Omega^*$, both in the zero-flux and in the thermal flux sectors.

\begin{figure}
	\centering
	\includegraphics[width=1.0\columnwidth]{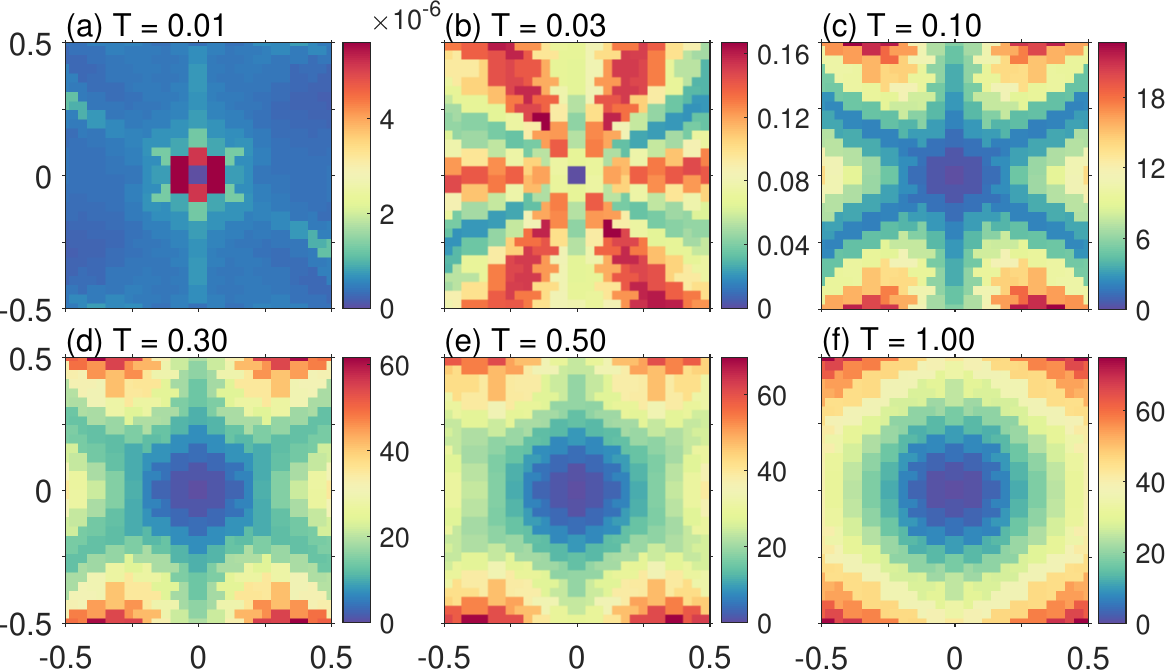}
	\caption{Case $v_s<v_F$: The temperature evolution of the sound attenuation coefficient $\alpha^{\|}_s(\vq)$  computed in the inhomogeneous flux sectors sampled by the strMC method.   The lattice size is $N_1=N_2=32$, and each data point is obtained by averaging over 200 flux realizations.
	 The phonon momentum ${\bf q}$ is in region $(q_x, q_y) \in [-0.5\pi, 0.5\pi]^2 $.  In the calculations, we set $v_s =0.1 v_F$ and the imaginary energy broadening  $\delta = 0.2$. Temperature is measured in units of the Kitaev interaction, $J_K$.}
	\label{MC_E2_32}
\end{figure}

\subsubsection{The sound attenuation coefficient: $v_s < v_F$ } \label{vslessvF}

We will first evaluate $ \alpha_s^{||} ({\bf q})$ assuming the zero-flux sector at all temperatures.   \reffg{kspace_attenuation_large} shows $ \alpha_s^{||} ({\bf q})$  
for various temperatures computed in  the system with $N_1= N_2 = 500$.
At all considered  temperatures,  the sound attenuation coefficient  displays the six-fold symmetry and   the pattern of the magnitude, which  agrees with the analytical results   of \cite{Ye2020}   (see \reffg{fig: anal_atten}  in App. \ref{App:att-analytical} for an explicit comparison of the angular positions of maximum and minimum  values). 

As the temperature increases, the overall intensity of the sound attenuation increases, which can be clearly seen from the increasing range of the colorbars. This is because the phase space of the Majorana fermion scattering in the ph-channel scales with temperature, thus at higher temperature, the intensity of the ph-channel scattering increases.
 It can also be observed that the area in the  phonon momentum space that actively contributes to the phonon's decay grows, which is a reflection that the scattering of the incident phonons with larger momentum  become  allowed by the kinematic constraints. When temperature reaches $T=1$ and goes beyond, the overall magnitude starts to decrease, as analyzed above in Sec.\ \ref{subsec: basic_picture}.

\begin{figure}
    \centering
    \includegraphics[width=1\columnwidth]{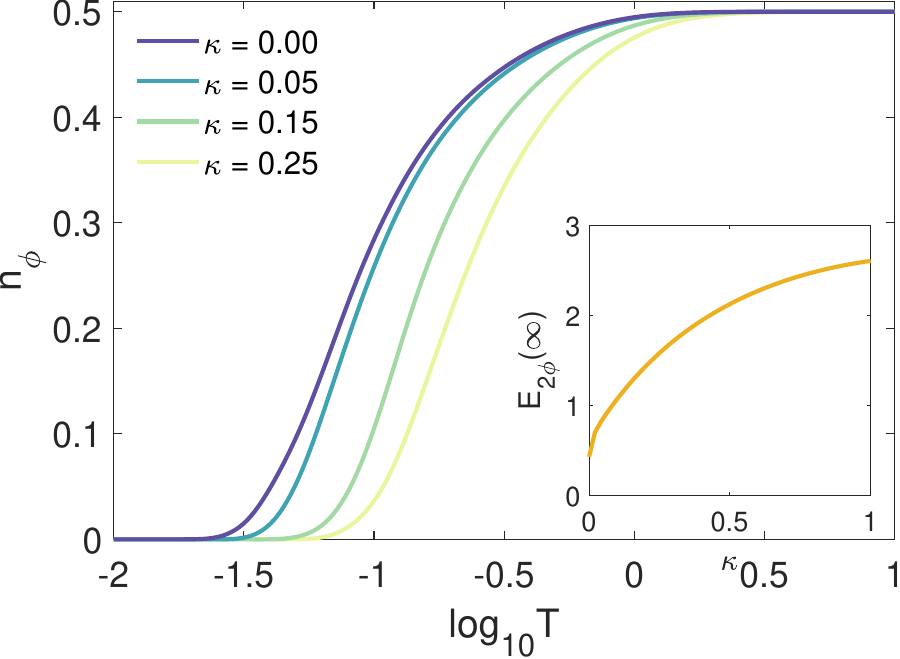}
    \caption{The temperature dependence of the flux density for various $\kappa$  obtained by the  strMC method on a lattice of size $N_1 = N_2 = 32$.  Each data point samples 50,000 flux configurations for each temperature. The inset shows the two-flux gap energy's dependence on $\kappa$ which is adapted from Ref. \cite{feng2020}. See the definition of the two-flux gap energy therein. The curve shows that the flux gap energy increases with $\kappa$. So, in the main panel, the flux onset temperature increases as $\kappa$ increases. Temperature and the two-flux gap energy  are measured in units of the Kitaev interaction, $J_K$.}
    \label{fig: flux_density}
\end{figure}

Above the flux onset temperature, the fluxes begin to proliferate. The evolution of the flux density  with increasing temperature obtained with   the strMC method is shown in \reffg{fig: flux_density}. For the case of unbroken time reversal symmetry, $\kappa=0$, the flux onset temperature is around $T^*\simeq10^{-1.5}\approx 0.03$.
Around $T_{\rm max}\simeq10^{0.5}\approx 3$, the flux density reaches its maximum  concentration, and the distribution of fluxes becomes completely random.

\reffg{MC_E2_32} shows  the sound attenuation coefficient $ \alpha_s^{||} ({\bf q})$ computed at various temperatures 
in the inhomogeneous flux sectors. $ \alpha_s^{||} ({\bf q})$  is obtained by  averaging over 200 flux configurations   at a  given temperature.  Already from
 a first glance comparison of  \reffg{kspace_attenuation_large}  and   \reffg{MC_E2_32}, we can see that 
  the attenuation coefficient pattern in the presence of fluxes  is very different from the pattern in the zero-flux sector shown in \reffg{kspace_attenuation_large}.  Indeed,  in the inhomogeneous flux sector, the peak and valley angular position are rotated compared  to those in the zero-flux sector leading to the appearance of the  star-like pattern. 
  Also, the overall magnitude  of the phonon decay increases with temperature remarkably faster  than in the zero-flux sector.

 There are three  factors that contribute into these modifications of the  sound attenuation coefficient. First, the overall magnitude  of $ \alpha_s^{||} ({\bf q})$ increases with temperature.  This happens due to the combined effect of the increasing with temperature 
 fermionic population  and  the appearance of additional low-energy fermionic modes associated with $Z_2$ flux sectors \cite{Lahtinen2011}.
 Second, the fermionic eigen functions are strongly modified in the presence of the  fluxes, which leads to the modifications of the coupling matrix elements of the phonon scattering (See \refeq{eq: Vq3}). Consequently, this change leads to the modification of the sound attenuation coefficient pattern. We also notice  that  in the temperature interval $T = 0.03\sim0.1$,  the star-like structure shows the 6-fold symmetry. The slight asymmetry of the pattern displayed near the flux onset temperature can be, perhaps, attributed to a finite-size effect, which gives rise to the inequivalence among the four topological sectors \cite{feng2020}. On the other hand, at  $T  \simeq 1$ an isotropic circular pattern begins to emerge, which is a combined effect of high population of fermion excitations and fully random flux configurations. 
Third,  as we discussed above, in the absence of the translational invariance, the kinematic constraint  related to the conservation of momentum is relaxed and  only the energy conservation constraint  remains. This allows for  more  fermionic modes to  contribute to the ph-channel scattering. This can be seen that at low temperatures the pattern of the inhomogeneous flux sectors has much larger active area in the phonon momentum space than that of the zero-flux sector.
To further analyze the  influence of the flux background on the pattern of the sound attenuation coefficient,  in App. \ref{App:att-fixeddensity} we calculate the average $\alpha_s(\vq)$ over uniformly sampled flux configurations  with fixed flux density.

\subsubsection{The sound attenuation coefficient: $v_s \geqslant v_F$ } \label{vsmorevF}

\begin{figure}
	\centering
	\includegraphics[width=1\columnwidth]{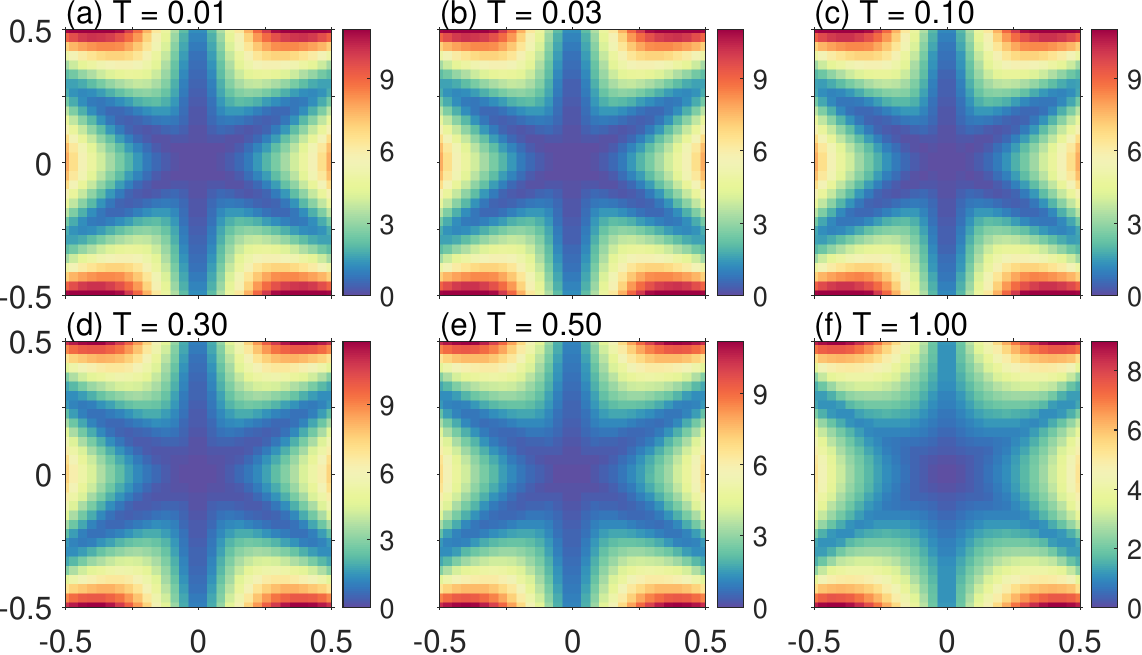}
     \caption{Case $v_s \geqslant v_F$: The temperature evolution of the sound attenuation coefficient  $ \alpha_s^{||} ({\bf q})$  computed in the zero-flux sector.  The diagonalization of the Majorana fermion Hamiltonian \refeq{eq:Hamfermion} is  done in the momentum space on the lattice with  $N_1= N_2 = 500$. The contributions from the pp- and ph-channels of  the scattering are summed up.
      The phonon momentum ${\bf q}$ belongs to the region $(q_x, q_y) \in [-0.5\pi, 0.5\pi]^2 $.  In the calculations, we set $v_s =1.1 v_F$ and the imaginary energy broadening  $\delta = 0.2$. Temperature is measured in units of the Kitaev interaction, $J_K$.
 }	\label{kspace_vsvf_phpp}
\end{figure}

We now turn to the analysis
of  the sound attenuation coefficient   for the case of $v_s \geqslant v_F$.
 \reffg{kspace_vsvf_phpp}  presents $\alpha^{||}_s(\vq)$  in the zero-flux sector obtained by the calculation in the momentum space of the finite-size system with $N_1= N_2 = 500$. Similarly to the case of $v_s < v_F$, at all considered temperatures, $\alpha^{||}_s(\vq)$ displays the  6-fold symmetry.
    However, the pattern of the decay intensity distribution  pattern is quite different from the   case of $v_s < v_F$ -- the peak's and valley's angular positions have been rotated compared to \reffg{kspace_attenuation_large}. Also the active phonon momentum region fills the whole region shown in the plot for all temperatures, while in the case of $v_s < v_F$, the active phonon momentum region grows with temperature. 
Another major difference lies in the  overall magnitude of the attenuation coefficient and its temperature dependence. In \reffg{kspace_vsvf_phpp}, as temperature increases, the overall magnitude of $\alpha^{||}_s(\vq)$ remains almost unchanged until $T = 1$, then begins to decrease, and the isotropic pattern appears. Recall, that in this case the dominant contribution into the phonon's decay comes from the pp-channel, in which the kinematic constraints can be always satisfied at  $v_s \geqslant v_F$. Thus,  already at the lowest temperature $T=0.05$,   
almost all fermionic modes are involved in the scattering unlike in the ph-channel where the scattering is limited by the available fermionic particle population.  Consequently, at  $T=0.05$  (see \reffg{kspace_vsvf_phpp} (a))  $ \alpha_s^{||} ({\bf q})$  already displays a star-like pattern. 
 With increasing temperature,
 the evolution of the pattern and the overall magnitude is mainly decided by the Fermi-Dirac function $n_F$, and  is in agreement
with the temperature evolution of the pp-channel scattering shown in \reffg{intensity}. This also verifies that even though both the ph- and pp-scattering channels  have been summed up in \reffg{kspace_vsvf_phpp}, the dominant one is  the pp-channel.

\begin{figure}
	\centering
	\includegraphics[width=1.0\columnwidth]{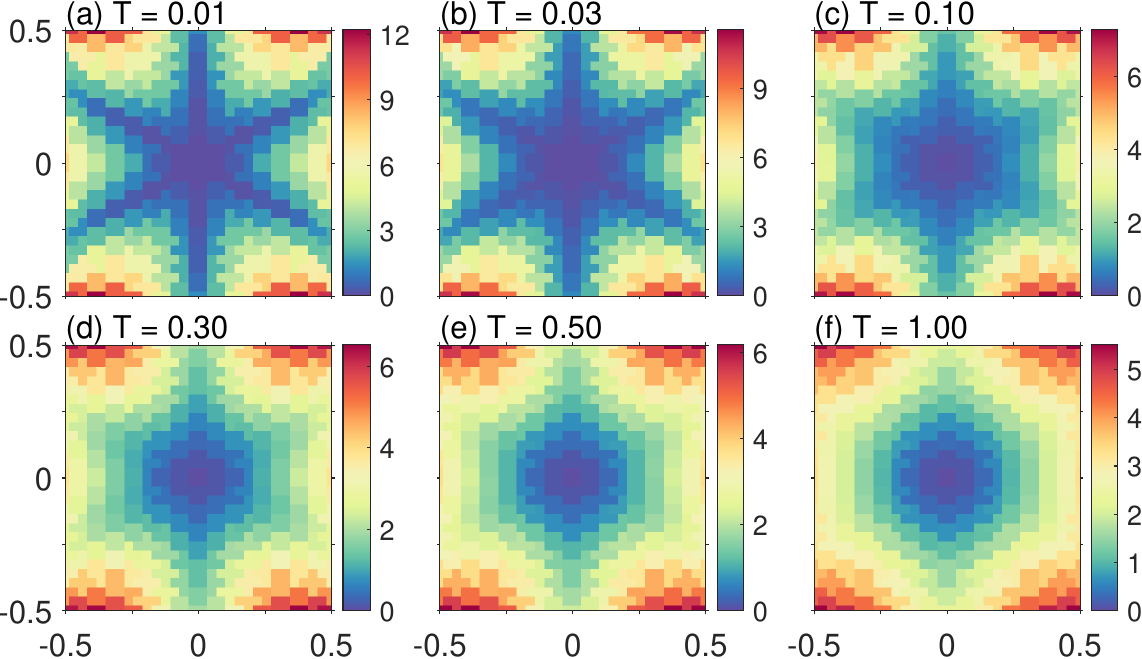}
     \caption{Case $v_s \geqslant v_F$:  The temperature evolution of the sound attenuation coefficient $\alpha^{\|}_s(\vq)$  computed in the inhomogeneous flux sectors sampled by the strMC method.   The lattice size is $N_1=N_2=32$, and each data point is obtained by averaging over 200 flux realizations.
	 The phonon momentum ${\bf q}$ is in region $(q_x, q_y) \in [-0.5\pi, 0.5\pi]^2 $.  The imaginary energy broadening is $\delta = 0.2$  and $v_s =1.1 v_F$. Temperature is measured in units of the Kitaev interaction, $J_K$.}	
     \label{MC_vsvf_phpp}
\end{figure}

\reffg{MC_vsvf_phpp}  shows    the sound attenuation for
 various temperatures computed in the inhomogeneous flux sectors.  At each temperature, $ \alpha_s^{||} ({\bf q})$
is  obtained by averaging over 200 random flux configurations sampled by the strMC method.  Similarly to the  behavior of sound attenuation in the zero-flux sector, the increasing of temperature doesn't change the pattern and the magnitude of the sound attenuation coefficient as dramatically as  in  the case of $v_s <v_F$ (see \reffg{MC_E2_32}). This can again be attributed to the difference in the underlying dynamics between the two scattering channels.  With increasing temperature and  the corresponding increase of the  flux density, the star-like pattern becomes blurry and smears into the isotropic pattern.


Similarly to the case of $v_s <v_F$, we also analyze the  influence of the flux background by disentangling  its effect from the  effect of  thermal population of the fermionic states. To this end, in App. \ref{App:att-fixeddensity} we  present the results of $\alpha_s(\vq)$ averaged over uniformly sampled flux configurations  with fixed flux density. \reffg{nflux_vsvf_phpp} shows that the star-like pattern remains almost unchanged when the density of fluxes increases.  \reffg{nflux_vsvf_phpp} also shows that the star-like pattern persists even at $n_\phi=0.5$,  when the flux configurations are totally random.

\section{Finite temperature effects on the phonon dynamics  without time reversal symmetry} \label{sec:V}

In this section, we will study the observable consequences of the  spin-lattice coupling 
when   time reversal symmetry  is broken by the three-spin interaction $\kappa$-term. 
Recall that this three-spin interaction term  changes  the  energetics  of  the  Kitaev model  (\ref{eq:Kmodel1}) by simultaneously gapping out the fermionic spectrum and introducing localized zero-energy Majorana modes in the presence of isolated fluxes \cite{Kitaev2006}.
The spin-lattice coupling Hamiltonian \refeq{Cmodel1}, however, does not contain a contribution from the three-spin interaction $\kappa$-term to the leading order (see Sec.\ \ref{sec:MFPh} for  more details). Also this term does not break the 6-fold rotation symmetry of the model. As a result, the decomposition of the spin-lattice coupling Hamiltonian into   $A_1$ and $E_2$ irreducible representations remains valid.
Therefore, the sound attenuation coefficient is modified only due to the changes in the fermion spectrum at low energy. 
Our calculations in App.\ \ref{sec: MCkap} indeed show that as $\kappa$ increases, the temperature evolution of the pattern of the sound attenuation $\alpha_s^{\|}(\vq)$ is similar to that of $\kappa=0$;  the only main difference is that the phonon decay starts at a higher temperature because of the increased fermionic gap energy as $\kappa$ increases.

Importantly, the $\kappa$-term breaks time-reversal and vertical mirror symmetries. What are the consequences to the phonon system? 
Here, we show that the phonon system acquires the Berry curvature induced by the $\kappa$-term due to the spin-lattice coupling, and study its evolution with temperature and the magnitude of $\kappa$.
The Berry curvature effect can be described by the Hall viscosity term in the phonon effective action
\cite{avron1995viscosity, barkeshli2012dissipationless,Ye2020}:
\begin{align}
    \mathcal{S}_{\mathrm{ph}}^{(a)}=\int \mathrm{d}^{2} x \mathrm{d} t\, \eta_{i j l k}^{(a)} \epsilon_{i j} \dot{\epsilon}_{i k},
\end{align}
where the viscosity tensor is anti-symmetric, i.e.\  $\eta^{(a)}_{ijlk}=-\eta^{(a)}_{lkij}$.  
It was also shown in Ref.~\cite{Ye2020} that by symmetry constraints, the Hall viscosity coefficient has  only one non-zero component $\eta_H$. 

Using the  linear response theory \cite{barkeshli2012dissipationless, Ye2020}, we can relate  $\eta_H$ to the off-diagonal component of the phonon polarization bubble:
\begin{align}
    \eta_H = \onefrac{q^2}\eta^{\perp\|} =  \onefrac{q^2\Omega l^d_a}\im\Pi^{\|\perp}(\vq,\Omega)|_{\Omega\to 0},\label{eq: eta_H}
\end{align}
where $l_a$ is the lattice constant.
Note that in the contribution to the Hall viscosity $\eta_H$, the off-diagonal component $\im\Pi^{\perp\|}(\vq,\Omega)$ must be anti-symmetric w.r.t.\ exchanging the polarization indices $\perp,\|$, which is required by the fact that the Hall viscosity is non-dissipative. Also, it is shown in Ref.~\cite{Ye2020} that the nonzero contribution to  $\im\Pi^{\perp\|}(\vq,\Omega)$ comes from the off-shell processes, in contrast to the dissipative on-shell processes involving the poles in $\im\Pi^{\perp\|}(\vq,\Omega)$.

In the following, we will discuss
the temperature evolution of the Hall viscosity coefficient $\eta_H$, 
focusing on understanding of the $Z_2$ fluxes effects.
We will show that  due to the distinct difference underlying the off-shell scattering processes compared with the on-shell scattering processes considered in previous sections, the temperature evolution of the Hall viscosity coefficient  is  very different from that of  the sound attenuation coefficient. We will also show that the magnitude of the Hall viscosity coefficient $\eta_H$  decreases rapidly with increasing density of the $Z_2$ fluxes.\\

\subsection{The Hall viscosity coefficient in the zero-flux sector}
\label{HVzeroflux}

We  start by analyzing the temperature evolution of the Hall viscosity coefficient in the zero-flux sector, when translation invariance allows to perform all the calculations in the momentum space. The  analytical derivation of the Hall viscosity coefficient at $T=0$ was performed in Ref.~\cite{Ye2020}, and the explicit expression of the off-diagonal component $\im\Pi^{\perp\|}(\mathbf{q}, i \Omega) $  that contributes to the Hall viscosity  is given by
\begin{widetext}
\begin{align}
    \Pi_{\mathrm{ph}}^{\perp \|}(\mathbf{q}, i \Omega) 
    =
    -\Tr\left[\hat{\lambda}_{\mathbf{q}, \mathbf{k}}^{\|} \mathcal{G}(\mathbf{k}, i \omega) \hat{\lambda}_{\mathbf{q}, \mathbf{k}}^{\perp} \mathcal{G}(\mathbf{k}-\mathbf{q}, i(\omega+\Omega))\right] = 
    &2 q^{2} \Omega \int \frac{\mathrm{d}^{2} k}{\mathcal{A}_{\mathrm{BZ}}} \frac{1}{\Omega^{2}+4\left|\vec{\mathcal{F}}_{\mathbf{k}}\right|^{2}} \frac{\Delta_{\mathbf{k}}}{\left|\vec{\mathcal{F}}_{\mathbf{k}}\right|}\left(\lambda_{x, \mathbf{k}, \mathbf{q}}^{\|} \lambda_{y, \mathbf{k}, \mathbf{q}}^{\perp}-\lambda_{y, \mathbf{k}, \mathbf{q}}^{\|} \lambda_{x, \mathbf{k}, \mathbf{q}}^{\perp}\right), \label{imPIperpparal}
\end{align}
\end{widetext}
where $\mathcal{G}(\mathbf{k}, i \omega)$ is  the Majorana fermion Green function, $
{\hat\lambda}_{\mathbf{q}, \mathbf{k}}$  denotes the Majorana fermion-phonon  coupling matrix,  which is expanded by Pauli matrices: $\hat{\lambda }= \sum_\alpha \lambda_\alpha \hat{\tau}^\alpha $, and
$\vec{\mathcal{F}}_{\bf k}=\{-\im f_{\bf k},-\re f_{\bf k},\Delta_{\bf k}\}$ with $f_{\bf k}=2J_K(1+2\cos \sqrt{3}k_x/2\, e^{i3k_y/2})$ and
$\Delta_{\bf k}=4\kappa \big(\sin \ve k\cdot\ve n_1-\sin \ve k\cdot\ve n_2+\sin \ve k\cdot (\ve n_1-\ve n_2)\big)$
[see \cite{Ye2020} for details of the derivation]. Note that $\vec{\mathcal{F}}_{\mathbf{k}}$ denotes the coefficients of decomposing the Majorana fermion Bloch Hamiltonian \refeq{eq:Hamfermion}  onto the basis of  the Pauli matrices. 

From \refeq{imPIperpparal} one  clearly sees that the contribution to $\Pi_{\mathrm{ph}}^{\mu\nu}(\qv,i\Omega)$ vanishes for $\mu=\nu$, and  that $\Pi_{\mathrm{ph}}^{\parallel\perp}(\mathbf{q}, i \Omega) =-\Pi_{\mathrm{ph}}^{\perp\parallel}(\mathbf{q}, i \Omega) $. Also, since  $\Pi_{\mathrm{ph}}^{\parallel\perp}(\mathbf{q}, i \Omega) \propto\Delta_\kv \propto \kappa $, 
it is nonzero only when time-reversal symmetry is broken. 
More formally, 
 the key  factor of yielding    $\Pi_{\mathrm{ph}}^{\parallel\perp}(\mathbf{q}, i \Omega) \neq 0$ lies in the fact that it is proportional  to  purely imaginary  and anti-symmetric quantity $\Tr\left[\hat{\tau}^{x} \hat{\tau}^{y} \hat{\tau}^{z}\right]=2i\epsilon^{xyz}$, where $\epsilon^{xyz}$ is Levi-Civita tensor.
 To get nonzero result, the three Pauli matrices must be a permutation of $\hat{\tau}^x, \hat{\tau}^y$ and $ \hat{\tau}^z$. Two of them are contributed from the coupling matrix $\hat{\lambda}$, which  are proportional to a combination of $\hat{\tau}^x$ and $\hat{\tau}^y$ matrices. 
  The third Pauli matrix comes from one of the propagators $\mathcal{G}(\mathbf{k}, i \omega)$, which comes with a coefficient  $\Delta_\vk$.  
  Consequently, $\im\Pi_{\mathrm{ph}}^{\parallel\perp}(\mathbf{q}, i \Omega) $ is obtained by taking the real part of the dynamical projection operators, whose equivalent real-space expressions  are shown in \refeq{eq:PPPP}.
  Since the real part of these operators is taken, the energy conservation constraint is absent, showing explicitly that the non-zero Hall viscosity originates from the off-shell processes, in which  all Majorana fermions contribute. Therefore, it is expected that the change in the fermionic population will not have any significant affect on the Hall viscosity at low temperatures (below any other energy scales in the problem, e.g. $\kappa,\, J_K$),
  in contrast to the on-shell processes contributing to the sound attenuation coefficient. In the high temperature region, where the particle and hole population becomes balanced, it is expected that the Hall viscosity coefficient will be suppressed to zero due to the Pauli exclusion principle.

\begin{figure}
	\centering
	\includegraphics[width=1.0\columnwidth]{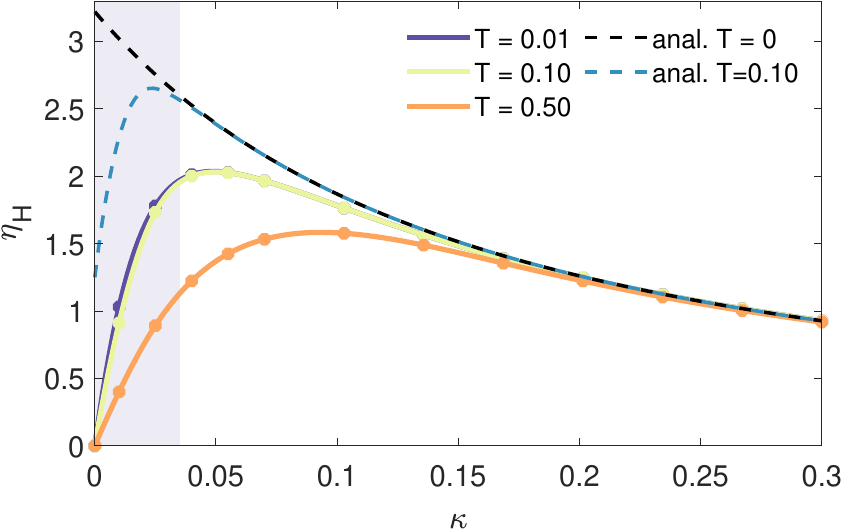}
	\caption{The Hall viscosity coefficient $\eta_H$ as of function of $\kappa$ computed for various temperatures in the zero-flux sector. The solid curves are obtained from the numerical evaluation of \refeq{eq:polarization}. 	The dashed curves are based on the analytical calculation of \refeq{imPIperpparal} in Ref.\cite{Ye2020}. 
    The lattice size used for the numerical calculation is $N_x=N_y=500$.  The phonon momentum and energy used in the calculation are ${\bf q} = [0, 0.1] \pi$ with $|\vq| = 0.36$ and $\Omega = 0.2$. The imaginary energy broadening $\delta=0.2$. The shaded area shows the region of small value of $\kappa$, where the deviation  of the numerical results  for  $\eta_H(T)$ from the zero temperature analytical result  $\eta_H(T=0)$ is the  most apparent. $T$, $\Omega$  and $\kappa$ are measured in units of  $J_K$. }
    \label{fig: eta_T_anal}
\end{figure}

This analysis is confirmed  by \reffg{fig: eta_T_kap} (a), where  the Hall viscosity coefficient $\eta_H$  computed in the zero-flux sector (purple line) at $\kappa = 0.05$ 
is plotted   as a function of temperature.
We can see 
 that below  $T=10^{-0.5}\sim 0.3$, $\eta_H$ remains almost unchanged. Above $10^{-0.5}$, the majority of Majorana fermions are excited and $\eta_H$ starts to decay to zero. The results for the non-zero flux densities will be analyzed in the next section.

The $\kappa$ dependence of the Hall viscosity $\eta_H$ in the zero-flux sector is shown in \reffg{fig: eta_T_anal}. The dashed curves are obtained from the analytical calculations \cite{Ye2020},  while the solid curves are obtained by the numerical evaluation of \refeq{eq:polarization}.  It shows that  for all values of $\kappa$, the Hall viscosity $\eta_H$ starts to decrease with temperature above $T=0.5$, which is consistent with the analysis above.
The comparison between the zero-temperature analytical curve (black dashed line)  and  the numerical curve computed at $T=0.01$ (blue solid line)  shows that  they agree well with each  when  $\kappa > 0.04$. However, when $\kappa<0.04$ (shaded region), the numerical  result for  $\eta_H$  deviates significantly from the $\eta_H$ computed analytically.  To clarify the origin of this discrepancy, we  plotted  another $\eta_H$-$\kappa$ curve at  $T=0.1$ (blue dashed line)  obtained from the analytical  calculation, where the contribution to $\eta_H$ from each fermionic mode is weighed by a factor $\tanh(\epsilon_k/T)$ calculated from the Fermi-Dirac distribution function, in the integration over the whole Brillouin zone. 
 So as temperature increases, the contribution from the  modes whose energies  are just above the  bulk gap introduced by  $\kappa$ will be suppressed by this factor. As shown in \reffg{fig: eta_T_anal},
the Hall viscosity $\eta_H$ in the shaded region decreases significantly at $T=0.1$. This indicates that the low-energy fermionic 
modes has a dominant contribution to $\eta_H$ at small values of  $\kappa$. 
Since  our numerical calculations are unable to capture the contribution from all low-energy modes due to the discreteness of energy levels in the finite size calculations,  our results show  significant 
decrease of $\eta_H$ when $
\kappa$ is very small. Therefore, the difference between the analytical results and the zero-temperature numerical results in the limit  of small $\kappa$ can be attributed to  the finite-size effects.
%
%
 The shaded region in  \reffg{fig: eta_T_anal},  \reffg{fig: eta_T_kap} (b)  and  \reffg{fig: eta_T_kap_MC} (b) shows where such deviation is the  most apparent.
  Note also that in the numerical calculation, we have to keep $\Omega$ small but finite, and in our calculations  we fix it as $\Omega=0.2$.
 In App. \ref{small_kap}, we  show the numerical verification that  for $\kappa=0.05$ the frequency $\Omega=0.2$  belongs to the region where  $\im\Pi^{\|\perp}(\mathbf{q}, i \Omega) $ is linear in  $\Omega$.

\subsection{The Hall viscosity coefficient in the inhomogeneous flux sectors}\label{HVinhomoflux}

\begin{figure}
	\centering
	\includegraphics[width=1.0\columnwidth]{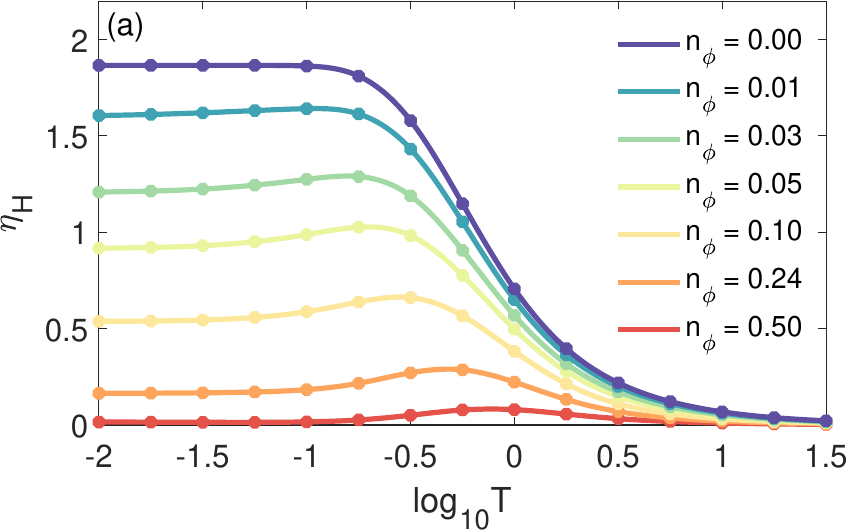}
    \includegraphics[width=1.0\columnwidth]{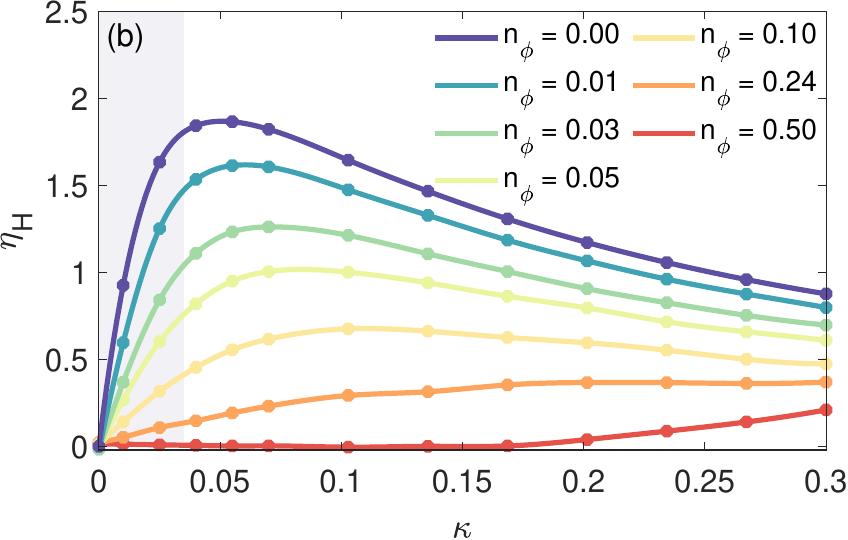}
	\caption{(a) The temperature dependence of $\eta_H$ for various flux densities computed at  $\kappa=0.05$. (b) The $\kappa$ dependence of $\eta_H$ for various flux densities at fixed $T=0.01$, which extends the $T=0$ result shown in \reffg{fig: eta_T_anal}. Each curve is obtained by taking the average over 200 flux configurations uniformly sampled for each flux density. $\im \Pi^{\|\perp}$ has been antisymmetrized w.r.t. exchanging the polarization indices.  In calculations we set 
	 ${\bf q} = [0, 0.1] \pi$ with $|\vq| = 0.36$ and $\Omega = 0.2$, which  are chosen based on the validation shown in \reffg{fig: eta_T_kap_valid}. The lattice size is $N_1 = N_2 = 32$. The imaginary energy broadening $\delta=0.2$. The shaded area in (b) shows the region of small value of $\kappa$, where the deviation  of the numerical results  for  $\eta_H(T)$ from the zero temperature analytical result  $\eta_H(T=0)$ is the  most apparent. $T$, $\Omega$  and $\kappa$ are measured in units of  $J_K$.
	}
	\label{fig: eta_T_kap}
\end{figure}


In this section, we  explore the effect of the $Z_2$ fluxes on the temperature evolution of the Hall viscosity coefficient. 
  In the presence of fluxes, the translational symmetry is broken and momentum is not a good quantum number. This means that the block diagonal matrix structure of  $\im \Pi^{\|\perp}(\mathbf{q}, i \Omega) $ will be mixed up by  non-zero off-block-diagonal entries. Consequently, the   proportionality of $\im \Pi^{\|\perp}(\mathbf{q}, i \Omega) $ to  $\Tr\left[\hat{\tau}^{\alpha} \hat{\tau}^{\beta} \hat{\tau}^{\gamma}\right]=2i\epsilon^{\alpha\beta\gamma}$, that is required for a nonzero Hall viscosity coefficient, is not guaranteed to be valid anymore.  Thus, we  expect that the Hall viscosity coefficient will be reduced.  Moreover, now $\im \Pi^{\|\perp} (\mathbf{q}, i \Omega)$ has  to be anti-symmetrized w.r.t. the exchange of two polarizations $\perp$ and $\parallel$.

\begin{figure}
	\centering
	\includegraphics[width=1.0\columnwidth]{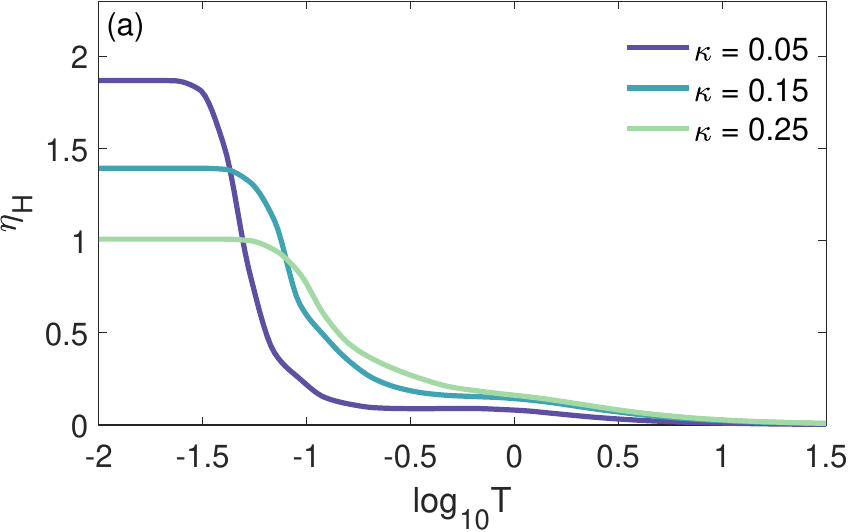}
    \includegraphics[width=1.0\columnwidth]{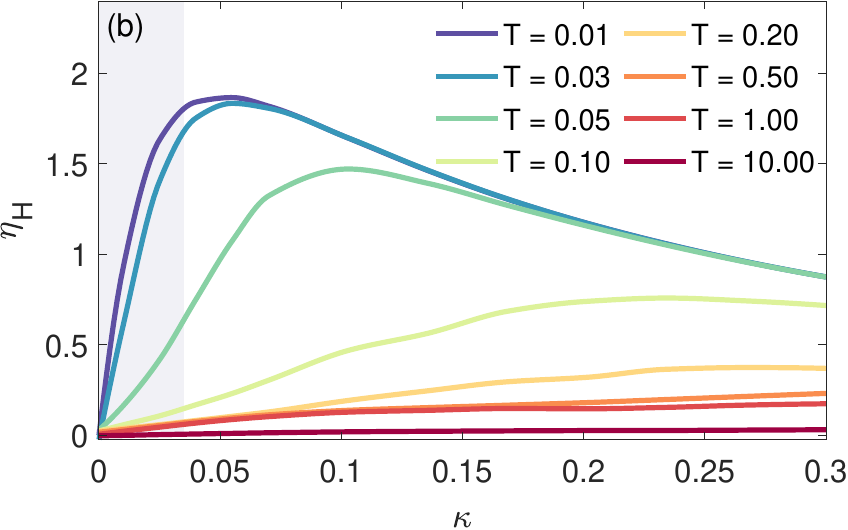}
	\caption{(a) The temperature dependence of  the Hall viscosity $\eta_H$  computed for various values of $\kappa$. (b) The dependence of $\eta_H$ on the strength of $\kappa$ computed for various temperatures. Each curve is obtained by the strMC method, which samples independent 100,000 flux configurations for each data point. $\im \Pi^{\|\perp}$ has been antisymmetrized w.r.t. exchanging the polarization indices. In calculations we set 
	 ${\bf q} = [0, 0.1] \pi$ with $|\vq| = 0.36$ and $\Omega = 0.2$, which  are chosen based on the validation shown in \reffg{fig: eta_T_kap_valid}.  The lattice size is $N_1 = N_2 = 32$. The imaginary energy broadening is $\delta=0.2$. The shaded area in (b) shows the region of small value of $\kappa$, where the deviation  of the numerical results  for  $\eta_H(T)$ from the zero temperature analytical result  $\eta_H(T=0)$ is the  most apparent.$T$, $\Omega$  and $\kappa$ are measured in units of  $J_K$.}
	\label{fig: eta_T_kap_MC}
\end{figure}


\reffg{fig: eta_T_kap} presents the Hall viscosity $\eta_H$  computed for various  flux densities. Each point is obtained by averaging over 200 randomly sampled flux configuration for a given flux density. Both $\eta_H$-$T$ curves  (\reffg{fig: eta_T_kap} (a)) and $\eta_H$-$\kappa$ curves  (\reffg{fig: eta_T_kap} (b)) show that the Hall viscosity decreases when the  density of fluxes increases. Also, this decrease is significant for a small increase of the flux density, which indicates that the Hall viscosity is sensitive to the matrix structure mixing introduced by the fluxes. Note that the time reversal symmetry is still broken which endows the elastic medium with nonzero Berry curvature, but due to the proliferation of the $Z_2$ fluxes, the Hall viscosity is reduced to basically zero.


 \reffg{fig: eta_T_kap} (b) shows $\eta_H$-$\kappa$ curves for various $n_\phi$ computed at $T=0.01$, and therefore it extends the  zero-flux sector results shown in \reffg{fig: eta_T_anal}. 
 In \reffg{fig: eta_T_kap}(b), the  $\eta_H$-$\kappa$ curves of small flux densities first increase in small $\kappa$ region and then decreases in large $\kappa$ region. The behaviour in the shaded region has been analyzed above. The behavior  in a region of  larger $\kappa$  in the presence of small flux densities is consistent with the zero flux case \cite{Ye2020}. Namely,
since the fermion energy gap $\Delta_\kappa$ increases with $\kappa$ \cite{Kitaev2006}, the contribution of the low-energy modes  into  the Hall viscosity is suppressed. Consequently, $\eta_H$ decreases with increasing $\kappa$.
 On the other hand, at large flux densities $n_\phi$, the  fermionic spectrum is significantly modified and part of the spectral weight near the van Hove singularity is shifted to lower energies \cite{Kao2021}. This may give rise to a 
small residual  Hall viscosity even at  relatively large $\kappa$.


 Next we study the evolution of the Hall viscosity coefficient in the presence of fluxes in a more realistic approach when  inhomogeneous flux configurations are  sampled by the strMC  method (see details in App.\ref{app: sMC}).  
 \reffg{fig: eta_T_kap_MC} shows the $\eta_H$-$\log T$ and $\eta_H$-$\kappa$ curves, obtained by the strMC algorithm which samples $100,000$ flux configurations for each point. In \reffg{fig: eta_T_kap_MC}(a), the Hall viscosity coefficients at the lowest temperature agree with the $\eta_H$-$\kappa$ curve at $T=0.01$ shown in \reffg{fig: eta_T_kap} (b). As temperature increases, flux density increases, so the $\eta_H$-$T$ curves start to decrease at $T= 10^{-1.5} \sim 10^{-1}$, which is earlier than  in \reffg{fig: eta_T_kap}(a). This again reflects the reduction the Hall viscosity coefficient by the $Z_2$ fluxes. \reffg{fig: eta_T_kap_MC} (a) also shows that the  $\eta_H$  for larger values of $\kappa$ starts to decrease at  higher temperatures.
This is because the flux gap energy $E_{2\phi}$ increases with $\kappa$ (see inset in \reffg{fig: flux_density}), so they proliferate at higher temperatures as $\kappa$ increases.

In \reffg{fig: eta_T_kap_MC} (b), we present the $\eta_H$-$\kappa$ curves computed at various temperatures. We can see that
the $\eta_H$-$\kappa$ curves display an overall downshift as temperature increases, which is mainly due to increased flux density. 
Also,  since at the same fixed temperature  the flux density in the  small $\kappa$ region is larger than that in the large $\kappa$ region (see \reffg{fig: flux_density}),  the Hall viscosity coefficient is suppressed stronger by the fluxes
 in the small $\kappa$ region than in the large $\kappa$ region. When temperature reaches $T=0.5$, where the flux density is close to saturation, the $\eta_H$-$\kappa$ curve becomes almost linear and diminishes to zero.

\section{Summary}\label{sec:sum}
In this paper, we proposed that the study of  the temperature evolution of the acoustic phonon dynamics
  can be used as potential probe of spin fractionalization in the Kitaev materials.
In our study we focused on two experimental observables -- the sound attenuation coefficient ($\alpha_s$) and the Hall viscosity coefficient ($\eta_H$). 
In particular, we  explored how  the sound attenuation and the Hall viscosity changes in the presence of the thermally excited $Z_2$ fluxes.
We showed that since the $Z_2$ fluxes  do not couple to the acoustic  phonons directly, their effect comes
mainly from providing the disorder potential for the  itinerant Majorana fermions and, thus, renormalizing their spectrum, and relaxing the kinematic constraints for the scattering processes. 

 We computed  the sound attenuation  and the Hall viscosity coefficients   by 
relating them to the imaginary part of the phonon polarization bubble.  To compute the bubble in the presence of the thermally excited $Z_2$ fluxes, when  the  translational  symmetry  for  the  Majorana  fermions  is  broken,  we derived
 a  microscopic low-energy effective spin-lattice coupling  Hamiltonian and formulated  a diagrammatic computation procedure in the mixed representation treating  the Majorana  fermions  in  the  real  space  and  the  acoustic  phonons  in the momentum space.


We found that the $Z_2$ fluxes can significantly change the  the sound attenuation, making it very different  from the one at the zero-flux sector.  We demonstrated that both  the angular dependence of $\alpha_s(\vq)$ and its magnitude show characteristic  changes when temperature is increasing and various inhomogenious flux sectors are being populated.
We showed that the strength of the flux effect on the phonon decay strongly depends on the ratio of the sound velocity and the Fermi velocity characterizing the low-energy Majorana fermions. Namely, we found that
when $v_s< v_F$, the thermal excitation of the $Z_2$ fluxes  increases the  overall  intensity  of  the sound attenuation and
changes the sound attenuation pattern  from  the flower-like pattern  in the zero-flux sector into the star-like pattern in the thermal flux sector. When $v_s> v_F$, both the overall intensity and the angular distribution of maxima and minima  of the sound attenuation coefficient are changing only slightly with temperature.
These differences reflect different scattering processes  contributing into the phonon decay in these two cases. They also show the combined effect of the proliferation of the $Z_2$ fluxes on the Majorana fermion-phonon coupling vertices, modified fermionic spectrum and relaxed kinematic conditions.

We found  that  the Hall viscosity coefficient, which is non-zero when time-reversal symmetry is broken, e.g., when external magnetic field is applied, decreases rapidly with increasing density of the $Z_2$ fluxes. Predominantly, this happens because in the presence of fluxes,  the block  diagonal  matrix  structure  of the imaginary part of the off-diagonal  component of the polarization bubble is  mixed  up  by the  non-zero  off-block-diagonal entries.  

To explore the phonon dynamics above the flux proliferation temperatures, we developed the  stratified Monte Carlo (strMC) algorithm to sample the flux configurations.  One of the main  advantages of this method  is that it has zero autocorrelation time, since the samples in the strMC method are independent. This  leads to faster convergence and less amount of samples, which is  particularly important in the calculations of the phonon dynamics.

 Finally, we note that our study was performed  for the pure Kitaev model. Of course, real Kitaev materials feature additional weak  time-reversal-invariant non-Kitaev interactions, which generally give rise to a wealth of nontrivial phases competing with the quantum spin liquids. 
While the Kitaev model looses its exact solubility in the presence of  additional  non-Kitaev terms,   we believe that the temperature evolution of the  sound attenuation  and the Hall viscosity will  remain similar to the one in the pure Kitaev model
as long these perturbations are small enough to leave the system in the Kitaev-like spin liquid phase.  In this case,  the effect  of the non-Kitaev terms on the dispersion of the Majorana fermions will be small, and their dominant effect  will be in adding  a dispersion to  otherwise localized dispersionless modes corresponding to the flux excitations. However, as long as these terms  will be small, the dispersion of the flux modes will be  small too, and the fluxes will be excited at more or less the same  temperature window as in the pure Kitaev model. Thus, the effect of thermal excitation of the fluxes on the Majorana fermions will remain similar and, consequently,  our results on the temperature evolution of the  sound attenuation  and the Hall viscosity will be robust  with respect to  a generic
time-reversal-invariant perturbation.

  \vspace*{0.3cm} 
\noindent{\it  Acknowledgments:} 
M.Y.\ and N.B.P.\ thank Fiona Burnell, Rafael Fernandes and Wen-Han Kao for valuable discussions. K.F.\ and N.B.P.\ were  supported  by  the  U.S. Department  of  Energy,  Office  of  Basic  Energy Sciences  under  Award  No. DE-SC0018056. 
   N.B.P.\ acknowledges the hospitality of Kavli Institute for Theoretical Physics and
 the National Science Foundation under Grant No. NSF PHY-1748958. 
 M.Y.\ was supported in part by the Gordon and Betty Moore Foundation through Grant GBMF8690 to UCSB and by the National Science Foundation under Grant No.\ NSF PHY-1748958. The authors thank the Minnesota Supercomputing Institute
 for providing computing resources, with which the numerical calculations in this paper were performed.


\appendix

\section{The Matsubara frequency summation in the evaluation of the polarization bubble \refeq{eq:PI}}
\label{app: eval_bubble}
We can evaluate the Matsubara frequency summation in using the standard residue method~\cite{AltlandBook}, and get the following expressions:
\bigskip
\begin{align}\label{eq:PPPP}
    P^{g\overline{g}}_{kl} &= T \sum_{i w_{n}} \frac{1}{i \omega_{n}-\epsilon_{k}} \frac{1}{\left(i \Omega_{m}-i w_{n}\right)+\epsilon_{l}} \nn
    \\
    &=\frac{n_{F}\left(\epsilon_{k}\right)-n_{F}\left(\epsilon_{l}\right)}{i \Omega_{m}-\epsilon_{k}+{\epsilon_l}},\nonumber\\
    P^{\bar{g} g}_{kl} &=T\sum_{i w_{n}}\frac{1}{i \omega_{n}+\epsilon_{k}} \frac{1}{\left(i\Omega_{m}-i \omega_{n}\right)-\epsilon_{l}} \nn
    \\
  &= \frac{n_{F}\left(-\epsilon_{k}\right)-n_{F}\left(-\epsilon_{l}\right)}{i \Omega_{m}+\epsilon_{k}-\epsilon_{l}} ,
  \\
    P^{\bar{g}\bar{g}}_{kl}  &=T\sum_{i w_{n}} \frac{1}{i \omega_{n}+\epsilon_{k}} \frac{1}{\left(i \Omega_{m}-i \omega_{n}\right)+\epsilon_{l}} \nn
    \\
    &= \frac{n_{F}\left(-\epsilon_{k}\right)-n_{F}\left(\epsilon_{l}\right)}{i \Omega_{m}+\epsilon_{k}+\epsilon_{l}} , \nn 
    \\
    P^{gg}_{kl}&=T\sum_{i w_{n}}\frac{1}{i \omega_{n}-\epsilon_{k}} \frac{1}{\left(i \Omega_{m}-i \omega_{n}\right)-\epsilon_{l}} \nn
    \\
    &=\frac{n_{F}\left(\epsilon_{k}\right)-n_{F}\left(-\epsilon_{l}\right)}{i \Omega_{m}-\epsilon_{k}-\epsilon_{l}}, \nn
\end{align}
where $P^{g\overline{g}}_{kl} ,  P^{\bar{g} g}_{kl} $ contribute to the particle-hole channel, $ P^{gg}_{kl}$ the particle-particle channel and $P^{\bar{g}\bar{g}}_{kl} $ the hole-hole channel.
 $n_F(\epsilon)$ is the Fermi-Dirac distribution function.

\begin{figure}[!h]
    \centering
    \includegraphics[width=0.5\columnwidth]{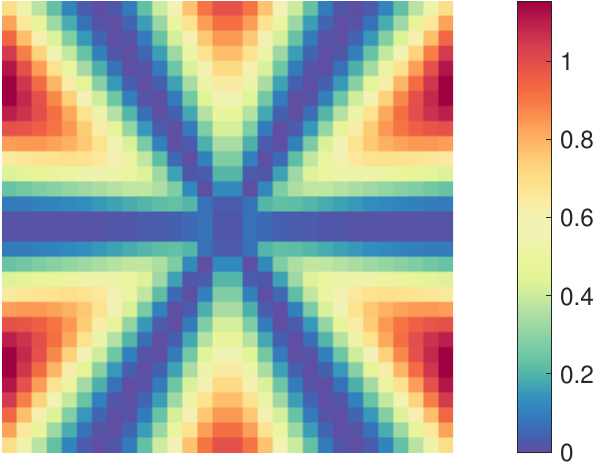}
    \caption{The plot of the attenuation coefficient $ \alpha_s^{||} ({\bf q})$  in the zero-flux sector computed from the analytical expression  \refeq{eq: anal_atten}.The result is shown in the $(q_x, q_y) \in [-0.5\pi, 0.5\pi]^2 $  region of the  Brillouin zone. To get a sense of how large is  this region,  one can compare the  boundaries of this region with the  positions of the high-symmetry points: $\Gamma = [0, 0], M \approx [0, 1.2]\pi, K\approx [0.66, 1.2]\pi, K' \approx [1.3, 0]\pi$.  In this plot, we set $v_s =0.1v_F$.}
    \label{fig: anal_atten}
\end{figure}

\begin{figure}[!h]
	\centering
	\includegraphics[width=1.0\columnwidth]{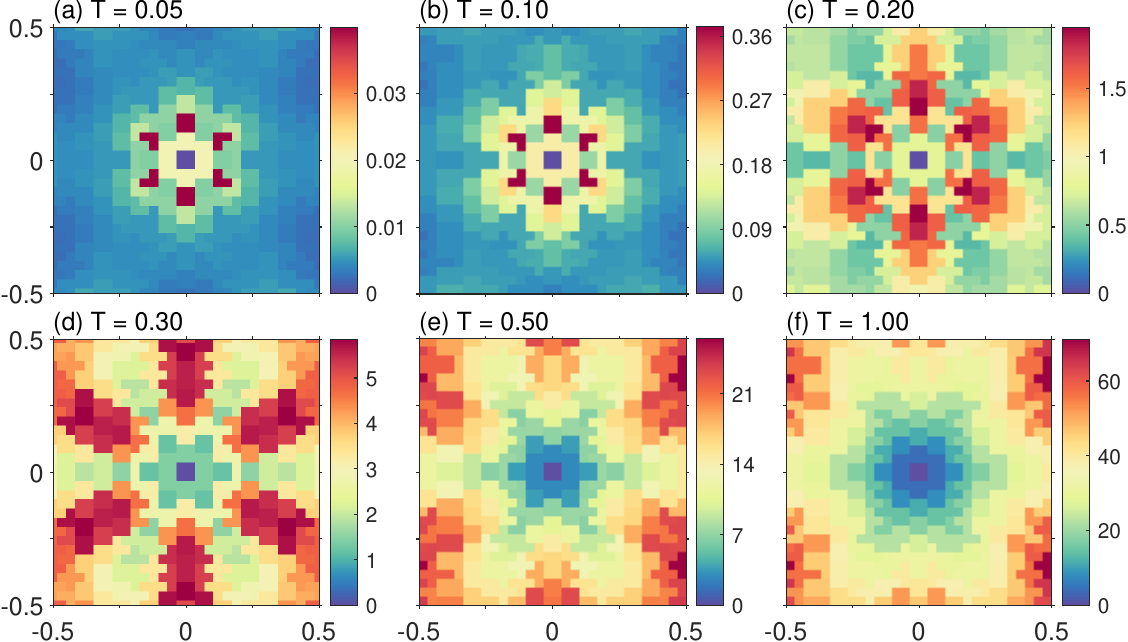}
    \caption{
    The temperature evolution of sound attenuation coefficient $ \alpha_s^{||} ({\bf q})$ for $v_s =0.1 v_F$, computed in the zero-flux sector with Majorana fermion eigenmodes obtained by the real space diagonalization of the Hamiltonian on $N_1= N_2 = 32$ lattice.
     The phonon momentum ${\bf q}$ is shown in region $(q_x, q_y) \in [-0.5\pi, 0.5\pi]^2 $.  The imaginary energy broadening $\delta = 0.2$. Temperature is  measured in units of  $J_K$.
    }
    \label{rspace_attenuation_32}
\end{figure}

\section{The  sound attenuation coefficient in the zero-flux sector}\label{App:att-analytical}

\reffg{fig: anal_atten} shows the magnitude of the sound attenuation coefficient $\alpha^{||}_s({\bf q})$ in the region of the Brillouin zone with
 $(q_x, q_y) \in ([-0.5, 0.5]\pi)^2$
computed in the momentum space according to the analytical expression $\alpha^{||}_s ({\bf q})\sim\frac{1}{q}  \operatorname{Im} \Pi_{\mathrm{ph}}^{\|\|}(\mathbf{q}, \Omega)$ with 
\begin{align}
  \operatorname{Im} \Pi_{\mathrm{ph}}^{\|\|}(\mathbf{q}, \Omega) \approx-\frac{36 \pi \lambda_{E_{2}}^{2} q|\Omega|}{v_{F}^{3} \mathcal{A}_{B Z}} T\left(1-\cos 6 \theta_{\vq}\right) \ln 2, \label{eq: anal_atten}
\end{align}
where only the leading order of the $q$ term is kept \cite{Ye2020}. Here, only the ph-channel of phonon decay contributes.
 The spatial distribution of $  \alpha_s^{||} ({\bf q})$ has the  six-fold rotational symmetry originated from  the $q(1-\cos(6\theta_\vq))$ angular dependence of $\operatorname{Im} \Pi_{\mathrm{ph}}^{\|\|}(\mathbf{q}, \Omega)$. We will use this result as a benchmark for the numerical calculations of the sound attenuation coefficient.

\begin{figure}[!t]
	\centering
	\includegraphics[width=1.0\columnwidth]{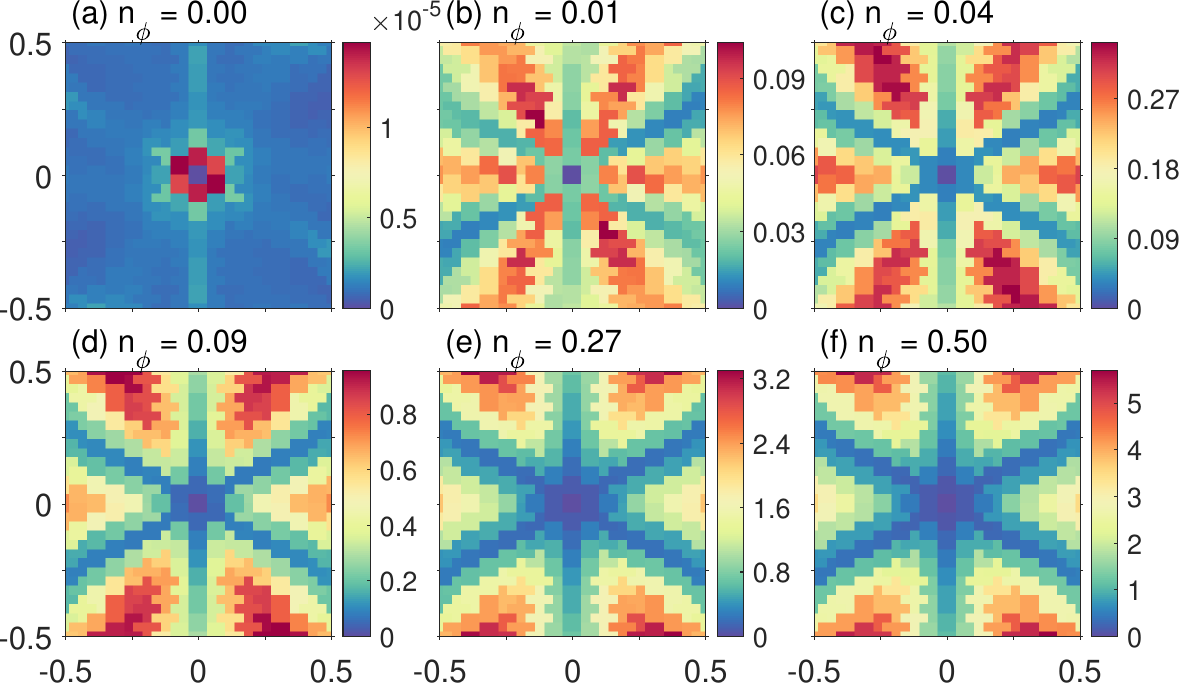}
	\caption{The sound  attenuation coefficient $\alpha^{\|}_s(\vq)$ for $v_s < v_F$ obtained by taking an average over 50 flux configurations uniformly sampled from each flux density at fixed temperature $T = 0.01$.  The choice of such a low temperature is made in order to show how the  fluxes modifies the low-energy fermionic wave function and its observable effect into the attenuation coefficient. It shows that the star-shape pattern seen in  \reffg{MC_E2_32}  is due to  the presence of the  flux background at finite temperatures. The phonon momentum ${\bf q}$ is shown in region $(q_x, q_y) \in [-0.5\pi, 0.5\pi]^2 $. The lattice size is $N_1=N_2=32$. The imaginary energy broadening $\delta=0.2$ and we set $v_s=0.1v_F$. Temperature is  measured in units of  $J_K$.}
	\label{nflux}
\end{figure}

The sound attenuation coefficient  in the  zero-flux sector can also be  computed from \refeq{eq: Pi_t}  by using the eigenmodes  of the Majorana fermions obtained by the diagonalization of the Hamiltonian in the real space instead of the momentum space. This is our approach  to the calculations  in the inhomogeneous flux sectors with broken translational symmetry, when the momentum is not  a good quantum number.  Because we sample different flux sectors with the Monte Carlo approach, simulations on large systems are costly  and we have to restrain ourselves to
simulations on  the $N_1= N_2 = 32$ lattice. 
 In order to separate the finite size effects  from the effect of fluxes, here we compare the sound attenuation computed in the zero-flux sector on  the $N_1= N_2 = 32$ lattice and shown in \reffg{rspace_attenuation_32} with \reffg{kspace_attenuation_large}, where diagonalization in the zero-flux sector was performed in the momentum space with $N_1= N_2 = 500$.
  This comparison tells that   apart from the finite size effects,  which are rather significant  in this case, \reffg{rspace_attenuation_32}  and  \reffg{kspace_attenuation_large} basically agree with each other. 
 At $T\geq 0.3$ (panels (d)-(f) on both plots), they show especially  good consistency both in terms of  the pattern of the angular distribution of the scattering probability and in the  magnitude.   However at low temperatures  (compare  (a)-(b)  panels of the corresponding plots),  
the finite size effects are stronger and, therefore, the discrepancies between the real space calculation and the momentum space calculations are more pronounced. 

\section{The sound attenuation coefficient $\alpha_s(\vq)$ averaged over uniformly sampled flux configurations  with fixed flux density.}
\label{App:att-fixeddensity}

\begin{figure}[!t]
	\centering
	\includegraphics[width=1.0\columnwidth]{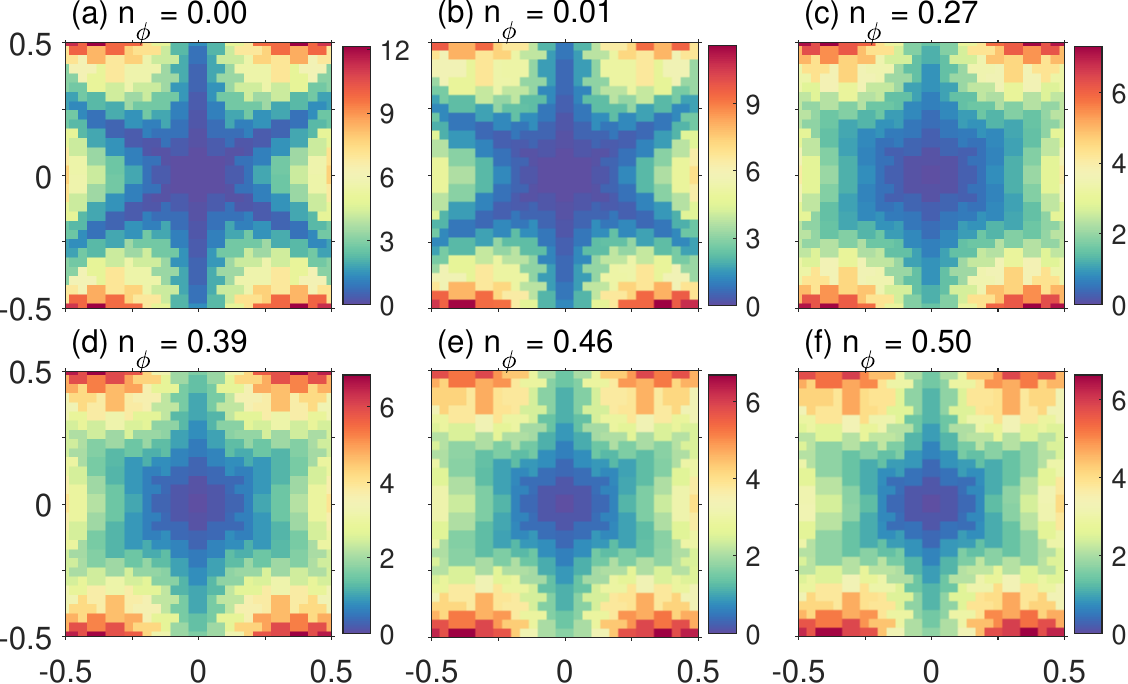}
     \caption{The sound attenuation coefficient $ \alpha_s^{||} ({\bf q})$ for  $v_s \geqslant v_F$    obtained by taking an average over 50 flux configurations uniformly sampled from each flux density at fixed temperature $T = 0.01$. The contributions from both the pp- and ph-channels of  scattering have been summed up. The choice of such a low temperature is made in order to show how the  fluxes modifies the low-energy fermionic wave function and its observable effect into the attenuation coefficient. 
      It shows that  increasing  flux density does not dramatically change the pattern  of $ \alpha_s^{||} ({\bf q})$ but only   smears it as the flux density increases.  This happens because of the relaxed  kinematic constraint in the presence of   fluxes. 
     As temperature increases, the fermionic modes are excited, and the result converges to the strMC  results in \reffg{MC_vsvf_phpp}. The phonon momentum ${\bf q}$ is shown in region $(q_x, q_y) \in [-0.5\pi, 0.5\pi]^2 $. The lattice size is $N_1=N_2=32$. The imaginary energy broadening $\delta=0.2$ and we set $v_s=1.1v_F$. Temperature is  measured in units of  $J_K$.}	\label{nflux_vsvf_phpp}
\end{figure}

To further analyze the  influence of the flux background on the pattern of the sound attenuation coefficient, here we present   $\alpha_s(\vq)$  computed at very low temperature  $T = 0.01$  and
averaged over uniformly sampled flux configurations. This allows to practically exclude the effect of thermal population of the fermionic states
 and analyze the effect of fluxes on the pattern  and the magnitude of the sound attenuation coefficient. 
 In  \reffg{nflux} and  \reffg{nflux_vsvf_phpp}, we  present  the results  computed  for $v_s=0.1\,v_F$ and  $v_s=1.1\,v_F$, respectively.

\section{The validation of the numerical approach  for the computation of  the Hall viscosity coefficient}
\label{small_kap}

\begin{figure}
	\centering
	\includegraphics[width=0.9\columnwidth]{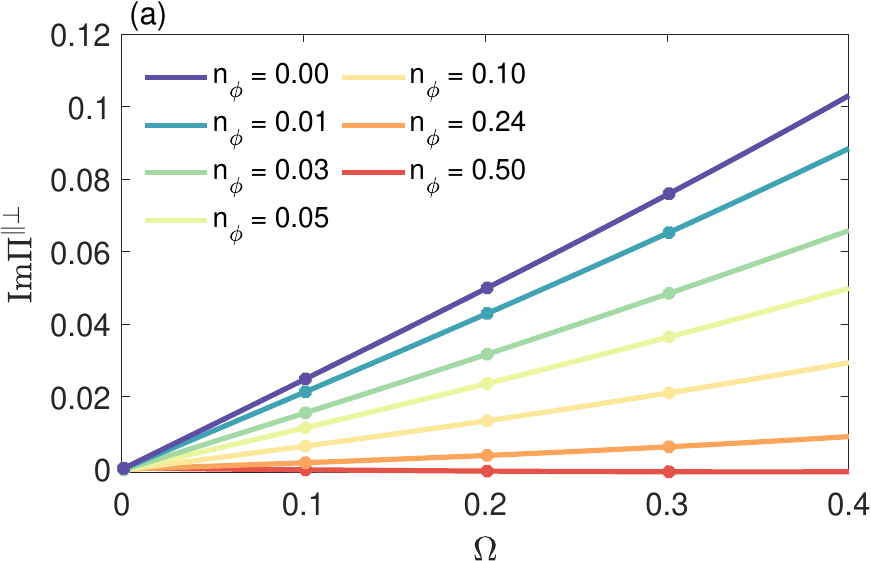}
	\includegraphics[width=0.9\columnwidth]{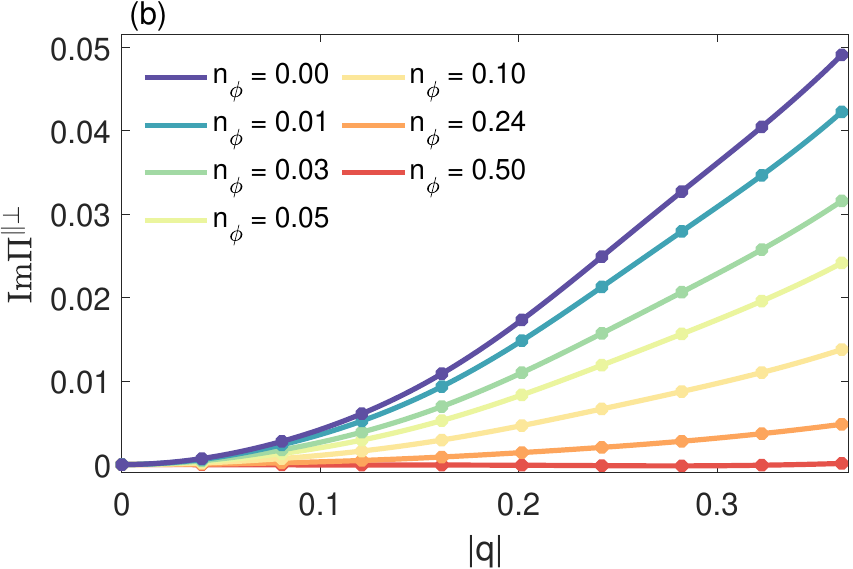}
	\caption{The imaginary part of the polarization bubble $\im\Pi^{\|\perp}(\vq, \Omega)$ plotted as a function of (a) phonon energy $\Omega$ and  (b) norm of phonon momentum $|\vq|$, for various flux densities.  Each curve is obtained by averaging over 200 flux configurations uniformly sampled for a given flux density. $\im \Pi^{\|\perp}$ has been antisymmetrized w.r.t. exchanging the polarization indices. Panel (a) shows the linear-$\Omega$ behavior of $\im \Pi^{\|\perp}$ in the low-energy region, and (b) shows  the quadratic-$|\vq|$ behavior in the small momentum region. The  values of $\Omega = 0.2$ and ${\bf q} = [0, 0.1] \pi$ with $|\vq| = 0.36$ are within the required region of the linear-$\Omega$ and the quadratic-$|\vq|$ behavior.
	 The temperature  is set to $T=0.01$ and we use $\kappa=0.05$. The lattice size is $N_1 = N_2 = 32$. The imaginary energy broadening is $\delta=0.2$. $T$ and $\Omega$ are measured in units of  $J_K$.}
	\label{fig: eta_T_kap_valid}
\end{figure}

\begin{figure}[!t]
	\centering
	\includegraphics[width=0.9\columnwidth]{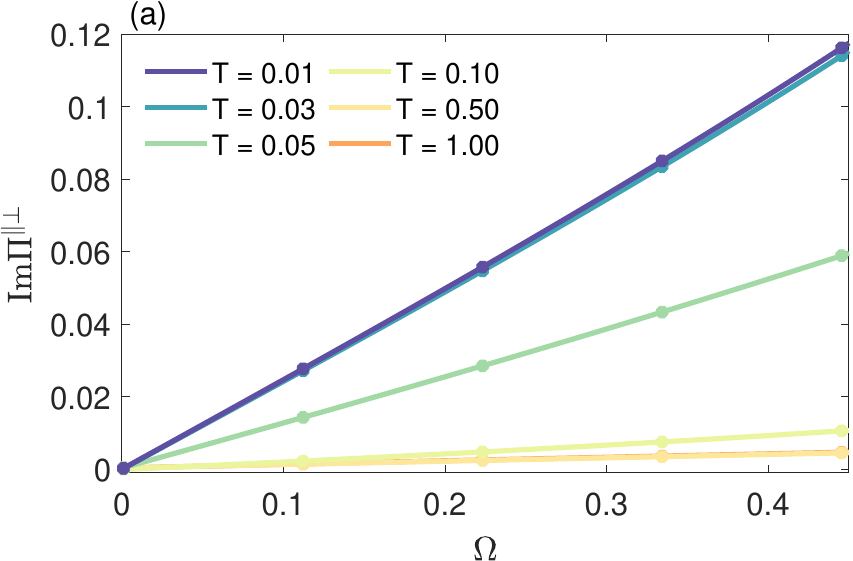}  
	\includegraphics[width=0.9\columnwidth]{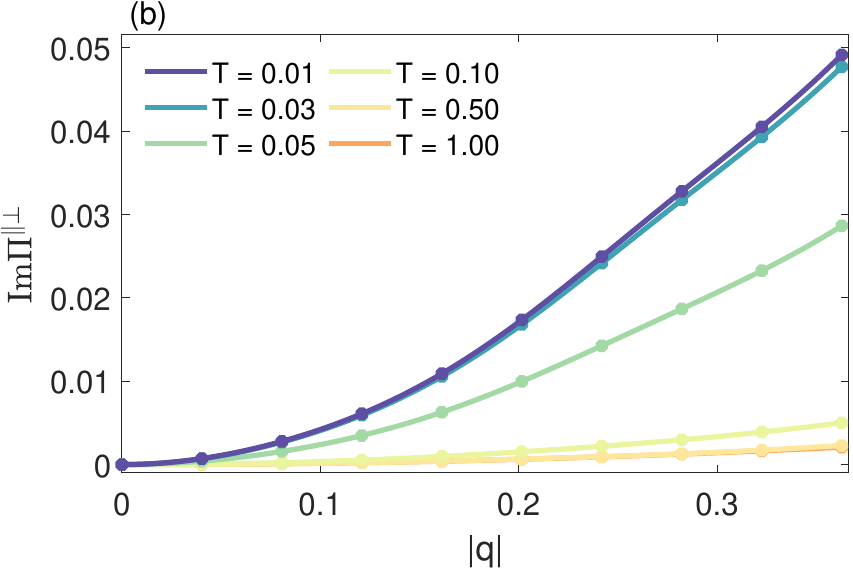}
	\caption{The imaginary part of the polarization bubble $\im\Pi^{\|\perp}(\vq, \Omega)$ plotted as a function of (a) phonon energy $\Omega$ and  (b) norm of the phonon momentum $|\vq|$, for different temperatures.
	  Each curve is obtained by averaging over 1000 flux configurations sampled by the strMC method. $\im \Pi^{\|\perp}$ has been antisymmetrized w.r.t. exchanging the polarization indices. Panel (a) shows the linear-$\Omega$ behavior of $\im \Pi^{\|\perp}$ in the low-energy region, and (b) shows  the quadratic-$|\vq|$ behavior in a small momentum region. The  values of $\Omega = 0.2$ and ${\bf q} = [0, 0.1] \pi$ with $|\vq| = 0.36$ are within the required region of the linear-$\Omega$ and the quadratic-$|\vq|$ behavior.
 The lattice size is $N_1 = N_2 = 32$. The imaginary energy broadening is $\delta=0.2$. $T$ and $\Omega$   are measured in units of  $J_K$.}
	\label{fig: MC_eta_T_kap_valid}
\end{figure}

The calculation of the Hall viscosity coefficient $\eta_H$ is based on \refeq{eq: eta_H}, which is  obtained 
 from the linear response theory through the Kubo formula.   This formula gives proper energy and momentum independent expression for
 $\eta_H$ if  the off-diagonal component of the polarization bubble $\im\Pi^{\|\perp}(\vq, \Omega)$  is  linear  in $\Omega$ and quadratic  
 in $|\vq|$.   The analytical  $T=0$ expression for $\im\Pi^{\|\perp}(\vq, \Omega)$ given by \refeq{imPIperpparal} clearly satisfies this requirement  \cite{Ye2020}. However, when we numerically compute  $\im\Pi^{\|\perp}(\vq, \Omega)$ according to \refeq{eq: Pi_expression}, we need to verify  that  $\im\Pi^{\|\perp}(\vq, \Omega)$ is indeed linear  in $\Omega$ and quadratic  
 in $|\vq|$, and confine our computations only to those $\Omega$ and $|\vq|$, for which it is satisfied.
 
  The validation results  for $\kappa=0.05$ are shown in \reffg{fig: eta_T_kap_valid},  where for  various  flux densities we plot (a) $\im\Pi^{\|\perp}$-$\Omega$ curves   in   a 
 low-energy region and  (b) $\im\Pi^{\|\perp}$-$|\vq|$ curves in small momentum region. Each curve is obtained by averaging over 200 flux configurations uniformly sampled for a given  flux  density. It shows that $\im\Pi^{\|\perp}$ is linearly dependent on $\Omega$ and quadratically dependent on $|\vq|$ for the flux densities shown in the figures. Our numerical calculation of $\eta_H$   for  various fixed flux densities  shown in \reffg{fig: eta_T_kap}  
uses $\Omega = 0.2$, which is within the linear-$\Omega$ region, and ${\bf q} = [0, 0.1] \pi$ with $|\vq| = 0.36$, which is small for the long wavelength approximation and within the quadratic-$|\vq|$ region.
 
 \reffg{fig: MC_eta_T_kap_valid} shows that the  polarization bubble $\im\Pi^{\|\perp}(\vq, \Omega)$  remains  linear  in $\Omega$ and quadratic   in $|\vq|$ in the  same low-energy and small-momentum  region when fluxes are sampled by the strMC method.
Each curve shown in \reffg{fig: MC_eta_T_kap_valid}   is  obtained by averaging over 1000 flux configurations sampled by strMC method. $\im \Pi^{\|\perp}$ has also been antisymmetrized w.r.t. exchanging the polarization indices. 
 


\section{The sound attenuation coefficient $\alpha_s(\vq)$ with time reversal symmetry breaking} \label{sec: MCkap}

\begin{figure}[!t]
    \centering
    \includegraphics[width=1.0\columnwidth]{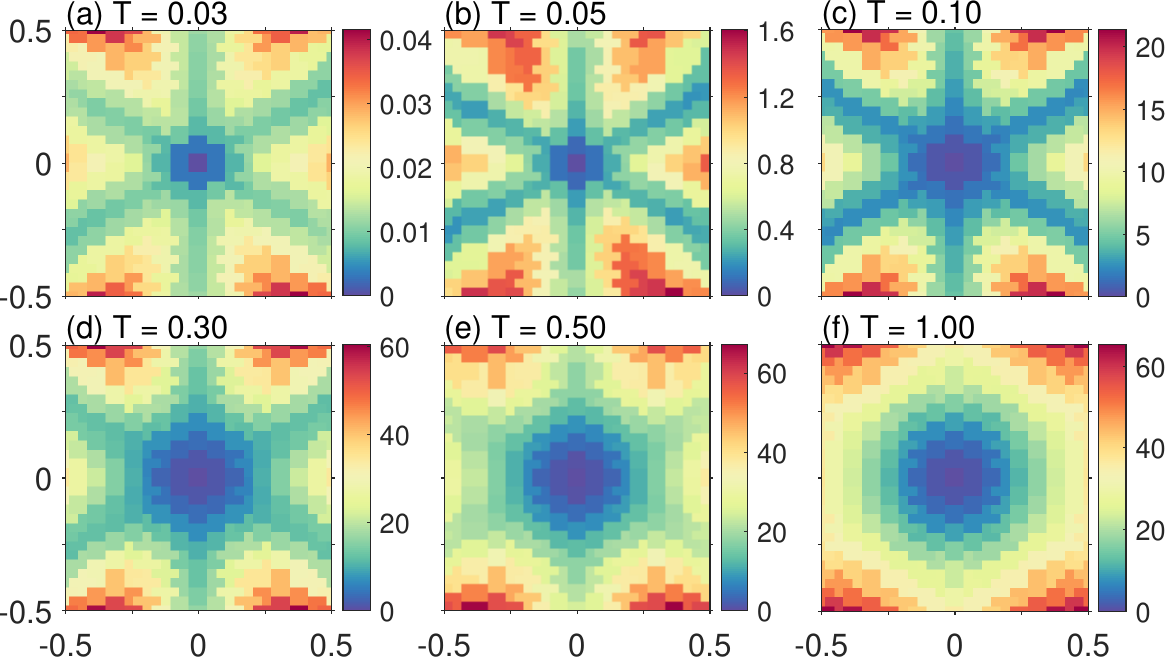}
    \caption{The temperature evolution of the sound attenuation coefficient $\alpha^{\|}_s(\vq)$  computed in inhomogeneous flux sectors sampled by the strMC method for the case $v_s<v_F$ and $\kappa=0.05$. The other parameters are the same as in \reffg{MC_E2_32}.}
    \label{fig: MC_atten_kap1}
\end{figure}

\begin{figure}{!t}
    \centering
    \includegraphics[width=1.0\columnwidth]{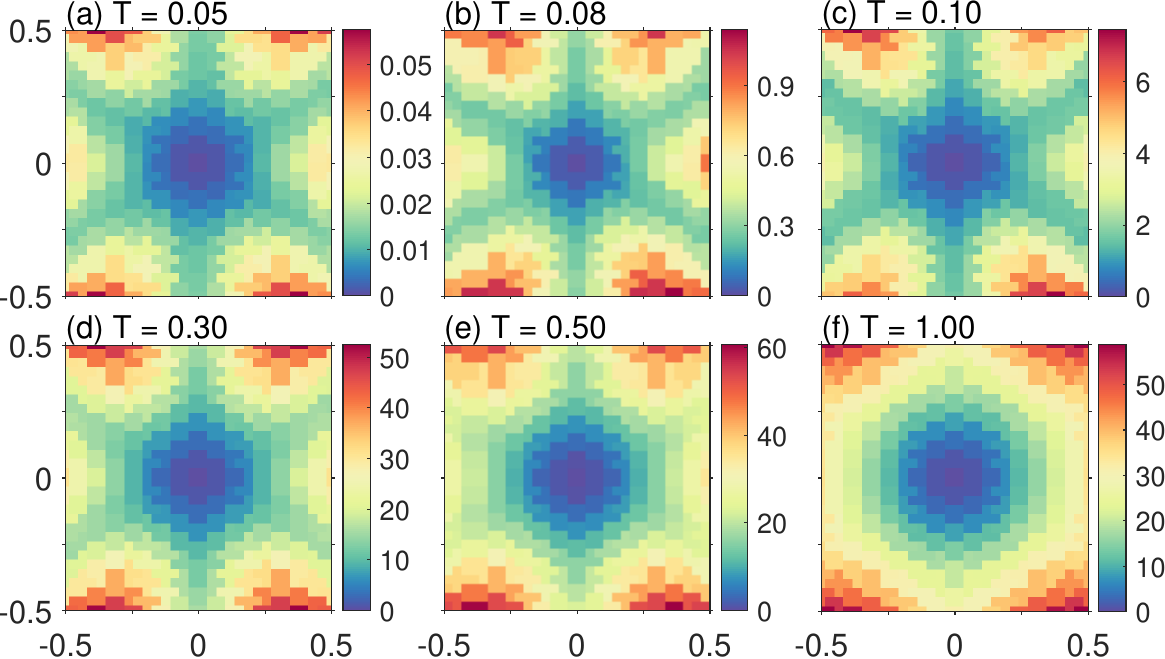}
    \caption{The temperature evolution of the sound attenuation coefficient $\alpha^{\|}_s(\vq)$  computed in  inhomogeneous flux sectors sampled by the strMC method for the case $v_s<v_F$ and  $\kappa=0.15$. The other parameters are the same as in \reffg{MC_E2_32}.}
    \label{fig: MC_atten_kap2}
\end{figure}

In order to verify that the sound attenuation coefficient when time reversal symmetry is broken, i.e. $\kappa>0$, is qualitatively the same as that when $\kappa=0$, we calculate $\alpha_s(\vq)$ for the case of $v_s < v_F$ for $\kappa=0.05$ in \reffg{fig: MC_atten_kap1} and $\kappa=0.15$ in \reffg{fig: MC_atten_kap2}. These plots are again obtained by strMC method and take the effects of the $Z_2$ fluxes into consideration. Comparing with \reffg{MC_E2_32}, we find indeed that the pattern of $\alpha_s(\vq)$ remains almost the same with increased $\kappa$. So the analysis in for $\kappa=0$ still applies. The only difference is in the overall magnitude of $\alpha_s(\vq)$ for temperatures below $T=0.1$: it decreases when $\kappa$ increases. This is because the fermionic energy gap $\Delta_\kappa$ increases with $\kappa$, so the low-energy modes starts to contribute to phonon's decay at higher temperature. For temperatures above $T=0.1$, the overall magnitude of $\alpha_s(\vq)$ remains almost unchanged. This is consistent with the fact that the overall fermionic band height remains almost unchanged for this range of $\kappa$ \cite{feng2020}.


\section{Stratified Monte Carlo method} \label{app: sMC}

The calculations that include the flux degrees of freedom at finite temperatures are done with the stratified Monte Carlo
method (strMC) \cite{rubinstein2016simulation,glasserman2013monte}. Different from a more generally applied Markov Chain Monte Carlo (MCMC) algorithm, this method is designed specifically for the Kitaev honeycomb model and is based on the flux energy model proposed in the recent work \cite{feng2020}. As will shown below, this method has several advantages over the MCMC method.

Stratified sampling \cite{rubinstein2016simulation} is a variance reduction technique in  the Monte Carlo simulations.  It consists of  dividing  the  sample  space  to  strata   and  then  estimating  the  weighted average of the yield from  each  stratum. 
 In this approach,  the samples are concentrated on the high probability regions, and the variance is thus reduced. It also offers an unbiased estimator to  thermodynamic quantities, and is in principle more efficient than the MCMC method, since the samples are independent from each other and have zero autocorrelation time. 

Specifically  to the Kitaev model, this method is designed to solve  the following problem. 
For a given observable $O(\{\phi_p\})$ whose value is determined by a certain flux configuration $\{\phi_p\}$ (with $p$ denoting the $p$-th plaquette and $\{\phi_p\}$ shorted as $\phi_p$ hereafter), the task is to find the thermodynamic expectation value of this observable over the ensemble of all flux configurations $\langle O \rangle_{\phi_p}$. 
We denote $I\equiv \langle O \rangle_{\phi_p}$, and
it can be generally computed as the following \cite{feng2020}:
\begin{align}\label{sum}
    I &= \sum_{\phi_p} p(\phi_p) O\left(\phi_{p}\right), \quad 
    p(\phi_p) = \onefrac{Z} e^{-\beta F(\phi_p)},
\end{align}
where $\beta=1/T$ and we assume $k_B=1$. $p(\phi_p)$ is the probability of the realization of a flux configuration $\phi_p$, which is decided by the free energy
$F(\phi_p) = E_\tx{flux}(\phi_p) - TS_F(\phi_p)$. Here, $E_\tx{flux}(\phi_p)=-\onefrac{2}\sum_k \epsilon_k(\phi_p)$, $S_F(\phi_p)=\sum_k \ln(1+e^{-\beta \epsilon_k}(\phi_p))$  and $Z = \sum_{\phi_p} e^{-\beta F(\phi_p)}$  is, respectively,  the flux energy, fermionic entropy and partition function of the system in  a given flux sector $\phi_p$. $\epsilon_k(\phi_p)$ denote the fermionic energy levels in the  flux configuration $\phi_p$. 
The central goal of this method is to generate a large set of sample flux configurations $\{\phi_p^i\}$ (where $i$ denotes the $i$-th sample), and use them to estimate the thermodynamic expectation $I$. The function evaluated from these samples  takes a form of sample average, which is called an estimator of $I$ and denoted as $\hat{I}$. In the following introduction to the strMC method, the focus will be on showing the way of generating the sample flux configurations $\{\phi_p^i\}$, the construction of the estimator $\hat{I}$, and the validation that $\hat{I}$ converges to $I$ in the large sample limit.

The summation in \refeq{sum}  is naturally stratified into the groups according to the number of fluxes denoted as $n$:
\begin{align}
    I& =
        \onefrac{Z}\sum_{n} \sum_{\phi_{p, n}} O\left(\phi_{p, n}\right) e^{-\beta F(\phi_{p,n})}  = \sum_{n} \frac{Z_{n}}{Z} \cdot I_{n},
\end{align}
where $\phi_{p, n}$ is a flux configuration restricted within an $n$-flux sector, and $I_n = \sum_{\phi_{p,n}} O\left(\phi_{p, n}\right) \frac{e^{-\beta F(\phi_{p,n})}}{Z_n}$, $Z_{n}=\sum_{\phi_{p,n}} e^{-\beta F(\phi_{p,n})}$ are respectively the thermal expectation and the partition function within the group denoted by $n$. It will be shown that the integral $I_{n}$ can be easily estimated. The number of samples allocated to each group is proportional to the group's probability $\frac{Z_n}{Z}$, so the sampling will be concentrated on the group with larger probability.  Then after obtaining the estimation of $I_n$ within each group, the final estimation of $I$ is obtained by a weighted sum over $I_n$, with weight proportional to the group's probability $\frac{Z_n}{Z}$.

\begin{figure*}
    \centering
    \includegraphics[width=1\textwidth]{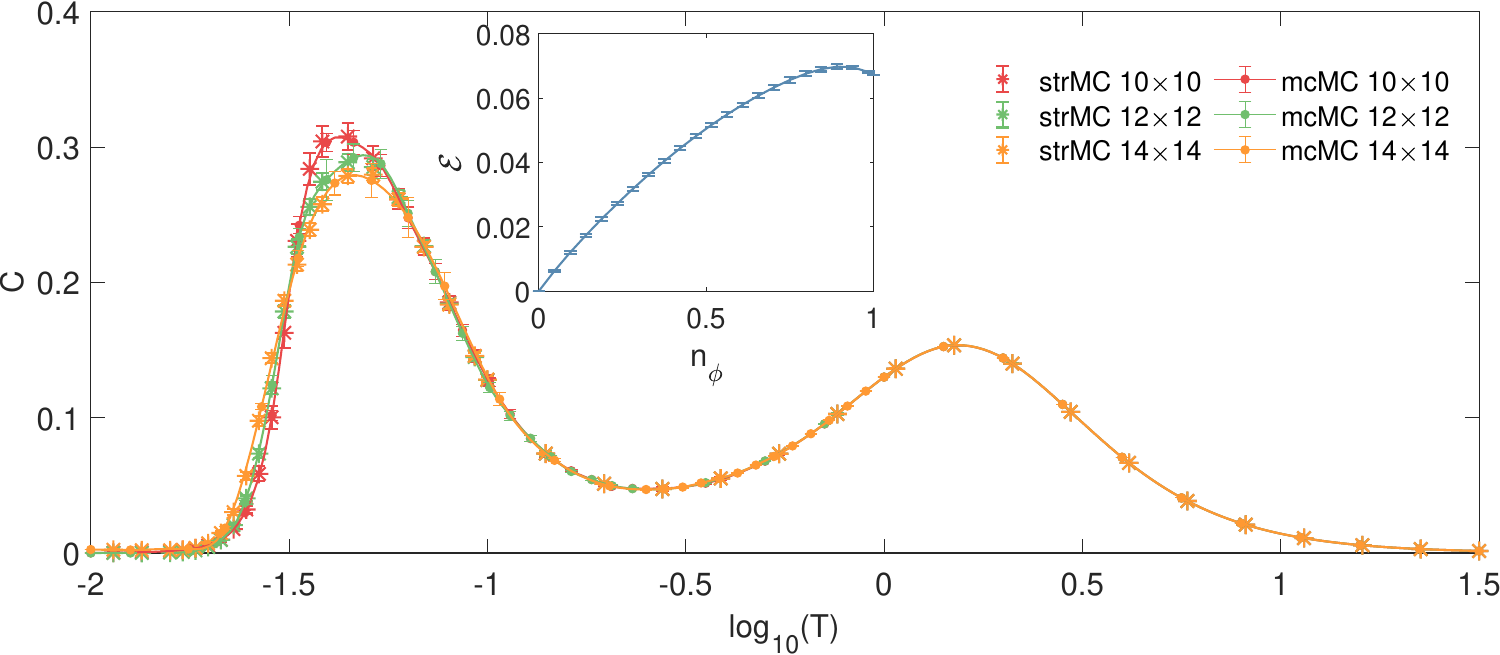}
    \caption{The temperature dependence of the specific heat of the Kitaev honeycomb model. The dashed curve is obtained by the MCMC algorithm, described in Ref.\ \cite{feng2020}, while the stars are obtained by the strMC algorithm. The total number of sample is 1,000,000 for each data point. The agreement between the two sets of results validates the strMC algorithm.  The inset shows the plot of the best-fit flux energy density $\mathcal{E}(n_\phi)$ as a function of the flux density $n_\phi$ for $\kappa=0$, which is adapted from Ref.\ \cite{feng2020}.  The flux energy $\bar{E}(n_\phi) = \mathcal{E}(n_\phi) N_p$, where $N_p$ is total plaquette number.  For the explicit expressions of the fit polynomial function and the result of other values of $\kappa$, see Ref.\ \cite{feng2020}.  Temperature and energies are measured  in units of $J_K$.}
    \label{fig: cmp_Cv}
\end{figure*}

To decide the number of samples allocated to each group, the probability of each group $\frac{Z_n}{Z}$ has to be known beforehand. To this end,  we find that this probability can be approximated by a phenomenological model, which describes the distribution  of the flux energies as a function of flux density. For sufficiently large lattices, this distribution is sharply peaked and is approximately independent of lattice size \cite{feng2020}.  Therefore, we can model flux thermodynamics by numerically fitting the flux energy with a polynomial curve. The resultant  best-fit flux energy as a function of flux density, so called  pseudo-potential energy (PPE) \cite{feng2020}, is shown in the inset of \reffg{fig: cmp_Cv}.
Basically, this best-fit flux energy $\bar{E}(n)$ and $W_n= \binom{N_p}{n}$,  the number of flux configurations for a fixed flux number with $N_p$ denoting the total number of plaquettes, decide the approximate probability of each group by $Z_n \approx W_n e^{-\beta\bar{E}(n)}$, 
and allocate the samples to each group accordingly.  
Note that this approximate weight $W_n e^{-\beta\bar{E}(n)}$ is only used to allocate the samples. When calculating the weighted average across the groups, this weight will be modified by a factor $\frac{\hat{Z}_n/W_n}{e^{-\beta \bar{E}(n)}}$, which will be shown next.


To estimate the thermal expectation value $I_n$, we rewrite it as the following:
\begin{align}
    I_n &  = \onefrac{Z_n/W_n}\sum_{\phi_{p, n}} O\left(\phi_{p, n}\right) \frac{e^{-\beta F\left(\phi_{p, n}\right)}}{W_{n}} .
\end{align}
Based on this expression, we can construct an estimator $\hat{I}_n$:

\begin{align}
    \hat{I}_n = \onefrac{\hat{Z}_n/W_n} \onefrac{N_n} \sum_{i=1}^{N_n} O\left(\phi_{p, n}^i\right) e^{-\beta F\left(\phi_{p, n}^i\right)}, \label{eq: E6}
\end{align}
where $\hat{Z}_n/W_n = \onefrac{N_n} \sum_i^{N_n} e^{-\beta F(\phi_{p, n}^i)}$ is relevant to the estimation of group partition function,
and $N_n$ is the  number of samples allocated to group $n$. $\phi_{p, n}^i$ denotes the i-th sample flux configuration that is generated from the uniform distribution within the flux sector $n$ and thus, are totally independent from each other. This is fundamentally different from the MCMC algorithm, where each running sample is probabilistically dependent on the previous samples. 

Now combining the estimations from all groups,  we can obtain an estimator of the integral $I$, denoted by $\hat{I}$:
\begin{align}
    \hat{I} &=  \sum_n \frac{\hat{Z}_n}{Z} \hat{I}_n = \onefrac{Z} \sum_n W_n \onefrac{N_n} \sum_{i=1}^{N_n} O\left(\phi_{p, n}^i\right) e^{-\beta F\left(\phi_{p, n}^i\right)}, \label{eq: estimator}
\end{align}
where the independently generated samples $\{\phi_{p,n}^i\}$ are used to estimate a thermodynamic average of a quantity $I$. But this expression is not convenient for translating to algorithm, so we rewrite it as
\begin{align}
    \hat{I} &= \onefrac{Z} \sum_n W_n e^{-\beta \bar{E}(n)} \frac{\hat{Z}_n/W_n}{e^{-\beta \bar{E}(n)}} \hat{I}_n, \label{eq: F12}
\end{align}
where the approximate weight $W_n e^{-\beta \bar{E}(n)}$, which determines the number of samples allocated into each group, has been modified by the factor $\frac{Z_n/W_n}{e^{-\beta \bar{E}(n)}}$, which is evaluated from the sampled data. 
To get a sense of this modification factor, note that  $\hat{Z}_n/W_n = \onefrac{N_n} \sum_i^{N_n} e^{-\beta F(\phi_{p, n}^i)}$ is the sample averaged Boltzmann weight of a flux sector, where $F(\phi_{p})$ is mainly decided by $E_\tx{flux}(\phi_p)$. So this modification factor is of the same order of $1$, which is understood as a tuning of the approximate weight.
After this tuning, the estimation $\hat{I}$ will be shown to be accurate, even though initially the approximate weight was used.

It remains to show that the estimator \refeq{eq: estimator} does not include systematic error, i.e.\ is unbiased. If an estimator of thermodynamic quantity is not constructed wisely, the sample average may not converge to the true thermal expectation value at large sample number limit.
This validation done by computing the expectation value of $\hat{I}$ on the random variable $\phi_{p, n}^i$ obeying uniform distribution $\tx p(\phi_{p,n}^i) =  \onefrac{W_n}$ within a flux sector $n$:
\begin{align}
    \mathbb{E}_{\phi_{p, n}^i}[\hat{I}] & = \onefrac{Z} \sum_n W_n \onefrac{N_n} \sum_{i=1}^{N_n} \mathbb{E}_{\phi_{p, n}^i} O\left(\phi_{p, n}^i\right) e^{-\beta F\left(\phi_{p, n}^i\right)} \nn
    \\
    & = \onefrac{Z} \sum_n W_n \onefrac{N_n} \sum_{i=1}^{N_n} \sum_{\phi^i_{p, n}}\onefrac{W_n} O\left(\phi_{p, n}^i\right) e^{-\beta F\left(\phi_{p, n}^i\right)} \nn
    \\
    & = \onefrac{Z} \sum_n \sum_{\phi_{p,n}} O\left(\phi_{p, n}\right) e^{-\beta F\left(\phi_{p, n}\right) } = I . \label{eq: expect}
\end{align}
So this estimator is indeed unbiased, which guarantees that as the number of samples $N$ goes to infinity, the sample average converge to the thermodynamic average \footnote{Careful readers may notice that the partition function $Z$ in $\hat{I}$ is also stochastic, 
since it is fine-tuned by $\frac{\hat{Z}_n/W_n}{e^{-\beta \bar{E}(n)}}$. Here, we have ignored this randomness in the expectation calculation \refeq{eq: expect}, since it can be shown by central limit theory that, in the large sample limit $N\to\infty$, $Z$ converges to the non-stochastic partition function \cite{casella2021statistical}. Therefore,  our approximation does not change the asymptotic behaviour of the estimator $\hat{I}$, which still converges to its thermodynamic expectation value.}.

In \reffg{fig: cmp_Cv} we plot the specific heat as a function of temperature (in a semi-logarithmic scale) calculated by both the MCMC and the strMC methods. \reffg{fig: cmp_Cv} displays a good agreement between the results from the two algorithms. Note that the low-temperature specific heat peak is rather sensitive to the thermal fluctuations as shown in Ref.\ \cite{feng2020}. Therefore, the good agreement between the two algorithms especially at the low-temperature peak validates the strMC algorithm we proposed.

 To conclude this Appendix, let us  underline the differences and similarities between
the MCMC and  the strMC  algorithms. In essence, both the  MCMC and the  strMC use the empirical probability  distributions.
The probability  distribution  in the  MCMC is given by
\begin{align}
    {p}_{\tx{MCMC}}(\phi_{p}) = \onefrac{N}\sum_{i=1}^N \delta(\phi_{p}- \phi_{p'}^i),
\end{align}
where $\{\phi_{p'}^i\}$ are sampled  according to their Boltzmann weights $p(\phi_{p'})\sim e^{-\beta F(\phi_{p'})}$, while the empirical distribution of the strMC is
\begin{align}
    {p}_{\tx{strMC}}(\phi_{p}) = \onefrac{Z} \sum_n W_n \onefrac{N_n} \sum_{i=1}^{N_n}e^{-\beta F(\phi_{p', n}^i)} \delta(\phi_{p}- \phi_{p',n}^i) ,
\end{align}
where $\{\phi_{p',n}^i\}$ are sampled from uniform distribution within flux sectors of a fixed flux number $n$, and then modulated by  their Boltzmann weights to get an unbiased estimation. 

An immediate advantage of the strMC over the MCMC is that it fundamentally reduces the autocorrelation time to zero, since the samples in the strMC are independent.  In  the recent works, where  the machine learning aided MCMC algorithms were applied to condensed matter systems \cite{huang2017accelerated,liu2017self,puente2020convolutional}, the deep neural networks have been employed to fit the system's free energy and to generate the MCMC updates that have an acceptance probability of nearly one. These works achieved a significant reduction on the autocorrelation time. The autocorrelation time in the strMC algorithm  is reduced to zero, leading to 
faster convergence and less amount of samples. This is  particularly important in the calculations of the phonon dynamics, where the main computation cost comes from the matrix production ${\bf B}^\dagger \tilde{{\Lambda}}^\mu_{\bf q} {\bf B}$ in \refeq{eq: Vq3}. 
Other advantages of the strMC method include the convenience in the implementation and parallelization, and the exemption from the annealing process (initial thermalization) and the local minima problems, which is common in the MCMC algorithms. 
Note, however,  that in the application of the  strMC in this paper, the division of the groups is specific to the Kitaev honeycomb model, where the flux number is naturally used to denote the group. For other models, the strMC method  needs to be specifically designed.

In a nutshell, in this strMC algorithm, the number of samples allocated to each group is based on the flux PPE model which is approximate, while the actual superposition weight across the groups is fine-tuned by the accurate probability distribution. Thus the estimation of the thermal expectation value is unbiased. This algorithm offers a good example of utilizing a pre-trained model to accelerate the Monte Carlo simulation, which is the idea shared by many machine learning algorithms \cite{huang2017accelerated,liu2017self, puente2020convolutional}. It is promising to combine this algorithm with the machine learning techniques to further improve the performance.

\bibliography{bib}
\end{document}